%% file: ms.tex
\documentclass[12pt,preprint]{aastex}

\usepackage{natbib}
\usepackage{epsfig}

\newcommand{\figref}[1]{Figure \ref{#1}}
\newcommand{\eqref}[1]{equation (\ref{#1})}

\newcommand{\sub}[1]{\ensuremath{_{\mbox{\scriptsize#1}}}}

\begin{document}
\slugcomment{To Appear in the Astrophysical Journal}
\title{Planet Shadows in Protoplanetary Disks. II: Observable Signatures}
\author{Hannah Jang-Condell}
\affil{Department of Astronomy, University of Maryland, 
College Park, MD 20742, U.S.A.}
\affil{NASA Goddard Space Flight Center, Greenbelt, MD 20771}
\email{hannah@astro.umd.edu}

\begin{abstract}
We calculate simulated images of disks perturbed by embedded small
planets.  These $10-50 \,M_{\earth}$ bodies represent the growing cores 
of giant planets.  We examine scattered light and thermal emission 
from these disks over a range of wavelengths, taking into account 
the wavelength-dependent opacity of dust in the disk.  
We also examine the effect of inclination on the observed 
perturbations.  
We find that the perturbations are best observed in the 
visible to mid-infrared.  Scattered light images reflect shadows 
produced at the surface of perturbed disks, while the infrared 
images follow thermal emission from the surface of the disk, 
showing cooled/heated material in the shadowed/brightened regions.  
At still longer wavelengths in the sub-millimeter, the 
perturbation fades as the disk becomes optically thin and 
surface features become overwhelmed by emission closer toward 
the midplane of the disk.  With the construction of 
telescopes such as TMT, GMT and ALMA due in the next decade, 
there is a real possibility of observing planets forming in disks 
in the optical and sub-millimeter.  However, having the 
angular resolution to observe 
the features in the mid-infrared will remain a challenge.  
\end{abstract}

\keywords{planetary systems: formation --- 
planetary systems: protoplanetary disks ---
radiative transfer}

\section{Introduction}
Giant planet formation remains an unsolved problem.  
The two leading theories are disk instability and core accretion. 
In disk instability, the protoplanetary disk gravitationally 
fragments into Jupiter-mass clumps which condense into planets.  
In core accretion, smaller bodies agglomerate into larger 
and larger bodies until they are massive enough to accrete 
a gaseous envelope.  There are theoretical arguments for and 
against both scenarios, but the aim here is to 
identify signatures of planet formation in order 
to provide direct observational evidence for one scenario or the 
other.  Simulated scattered light signatures of planet formation 
by disk instability have been discussed elsewhere 
\citep{2007HJCBoss}, so the focus of this paper is on signatures of 
core accretion.  

Giant planets forming by core accretion need to have cores of 
10-20 M$_{\earth}$ to be massive enough to 
accrete a gaseous envelope \citep{2005Hubickyj_etal}.  
For reference, Jupiter is approximately 300 M$_{\earth}$.
This predicts that sizeable planet embryos form before 
circumstellar gas disks dissipate.  
These disks are typically modeled as passive accretion disks, 
meaning that the dominant heating source is stellar irradiaion, 
although there still is sufficient viscous dissipation 
for the disk to continue accretion at a slow rate onto the 
star ($\lesssim 10^{-7}\,M_{\sun}\mbox{yr}^{-1}$) and 
for a modest amount of heating at the disk midplane.  
These disks are optically thick and gas-dominated.  
Their temperature structure is strongly dependent on heating from
stellar irradiation, particularly on the angle of incidence of the
starlight on the disk surface
\citep{CG, calvet, vertstruct, dalessio2, dalessio3}.

We examine planets forming at distances of $1-8$ AU from the stars.  
Interior to 1 AU, 
the innermost regions of protoplanetary disks are expected to become 
optically thin because of the sublimation of dust grains,
creating a hot inner rim at $\sim 0.1-0.5$ AU 
\citep{DDN,2005IsellaNatta,2006DAlessio_etal}.  
This is supported by observations of T Tauri stars in the near- to mid-IR 
\citep{2009Eisner_etal,eisner2005,2003Muzerolle_etal}.  
Although larger radii for inner holes in protoplanetary disks are observed, 
this is typically for higher mass Herbig Ae/Be stars, which are both 
hotter and more luminous than T Tauri stars, resulting in the 
evaporation of dust grains at larger distances from the star.  
The typical inner disk radius for T Tauri stars is below 
0.5 AU, which is the minimum disk distance modeled in this study.  
For this reason, the emission from the inner disk rim can be treated 
as a separate problem from the study presented in this paper.  

Other work on simulated images of growing planets focuses 
on Jovian-mass planets rapidly accreting material 
from their surroundings disks.  These planets are large 
enough to create large-scale structures in disks 
such as gaps or spiral density waves, and the 
planets themselves are observable because they are heated 
by accretion \citep[e.g.~][]{2008Wolf,2006KlahrKley}.  
Planet formation has already completed and it is difficult 
to determine the mechanism by which it occured.  
By contrast, the study presented in this paper focuses on 
planets which are too small 
to open gaps or accrete significant amounts of gas ($10-50\,M_{\earth}$).  

The planet-disk models adopted here are presented in \citet{HJC_model}
and summarized in this section.  
The planet is predicted to gravitationally compress the disk in the vertical 
direction, creating a shadow paired with a bright spot, leading 
to temperature variations.  We calculate both scattered light 
and thermal emission from the perturbed disk for planets of mass 
$10-50\,M_{\earth}$ at distances of $1-8$ AU from their host stars.  

In \S\ref{model}, we describe the procedure we use for modeling 
radiative transfer in the disk.  In \S\ref{results} we show the 
results of the modeling for the full suite of planet-disk models 
over a range of wavelengths.  In \S\ref{discussion} we discuss 
the potential observability of the planet perturbations we have 
modeled.  In \S\ref{conclusion} we present our conclusions.

\section{Model Description}\label{model}

\subsection{Disk Model: Thermal Structure}

The model for thermal perturbations in the vicinity of 
an embedded protoplanet are presented in \citet{HJC_model}. 
The stellar parameters are 
mass $M_*=1\,M_{\sun}$, 
radius $R_*=2.6\,R_{\sun}$, and 
effective temperature $T\sub{eff}=4280$ K\@.
The radius and effective temperature correspond to a stellar 
age of 1 Myr \citep{siess_etal}.  
We assume a viscosity parameter of $\alpha=0.01$
and accretion rate $\dot{M}= 10^{-8}\,M_{\sun}\mbox{yr}^{-1}$, 
which are typical of T Tauri stars \citep{hartmann,vertstruct}.  

The model for the disk is a based on an 
accretion disk model 
with viscosity parametrized by a dimensionless constant 
$\alpha$, so that the viscosity is $\nu = \alpha c_s H$
where $c_s \equiv \sqrt{kT_0/\bar{m}}$ is the thermal sound speed 
at the midplane of the disk 
and $H \equiv c_s/\Omega$ is the pressure scale height
\citep{shaksun}.  
Here, 
$k$ is the Boltzmann constant, $T_0$ is the midplane temperature, 
$\bar{m}$ is the mean molecular weight of molecular hydrogen, 
and $\Omega$ is the Keplerian angular velocity.  
For the low accretion rate adopted and 
the distances in the disk in which we are interested, 
stellar irradiation heating dominates, though we include 
viscous heating for completeness.

The amount of heating from stellar irradiation at the surface 
depends on the angle of incidence of the stellar rays.  
Based on an analytic solution to the 1-D plane-parallel case, 
we calculate surface heating semi-analytically in a 1+2D manner, 
where we numerically integrate heating contributions over the 
shape of the disk surface.  
We include the differential rotation of the disk by calculating 
the rate of radiative 
cooling and heating as disk material enters or exits 
the shadowed or brightened regions.  

We assume vertical hydrostatic equilibrium in the disk, 
including contributions to the vertical gravity 
from both the star and the planet.  
We treat the planet embedded in the disk as a point mass, whose 
gravitational potential 
compresses the disk in the vertical direction, creating a dimple at the 
surface.  
The dimple at the surface results in a shadow at the 
planet's position relative to the unperturbed disk,
while the outer edge is brightened.  

The radiative transfer calculations rely on a separation of the 
radiation into two regimes: the short wavelength spectrum of the 
stellar emission, and the long wavelength spectrum of the 
thermal emission of the disk material.  
We use the opacities from \citet{dalessio3} using a dust model with 
parameters $a\sub{max} = 1\,\mbox{mm}$, $T=300$ K, and $p = 3.5$, 
assuming that the dust opacities are constant throughout 
the disk.  The values for the opacities (in $\mbox{cm}^2 \mbox{g}^{-1}$)
are as follows: the Rosseland mean opacity is $\chi_R=1.91$, 
the Planck mean opacity integrated over the disk spectrum (300 K) is $\kappa_P=0.992$,
and the Planck mean opacities integrated over the stellar spectrum (4000 K) are 
$\kappa_P^*=1.31$ for absorption alone and $\chi_P^*=5.86$ for absorption plus scattering.  
The absorption fraction is then 
$\alpha\sub{abs} = \kappa_P^*/\chi_P^*$, 
while the scattered fraction is $\sigma = 1-\alpha\sub{abs}$.  
The Rosseland mean opacity is used to calculate the photosphere of the 
disk, and $\chi_P^*$ is used to calculate the surface of the disk.  
We assume that the dust is well-mixed with the gas 
and constant throughout the disk.

The initial conditions vary slightly in the overlap regions 
between simulations as the 
size and position of the simulation box vary.  The sizes of the 
boxes are scaled to the Hill radius, 
\begin{equation}
r\sub{Hill} = \left( \frac{m_p}{3 M_*} \right)^{1/3} a.  
\end{equation}
Thus, the boundary conditions depend on the mass and the distance 
of the planet.  In order to make a fair comparison between 
the various planet models, we assemble a composite disk from the 
initial conditions of the largest boxes at each distance from the star.  
We interpolate the densities and temperatures smoothly across the 
overlap regions.  The magnitude of the variations in temperature and 
density should roughly scale, so we normalize the temperatures 
and densities of each planet-disk model to this composite set of 
initial conditions.

\subsection{Opacities}

In order to calculate simulated images of the disk, we need a good 
model for the opacities.  Although mean opacities were good enough for 
calculating the thermal structure of the disk, to calculate 
images of the disk, we need the frequency-dependent 
absorption and total extinction coefficients for the dust model.  

The opacities were calculated using a Mie scattering code developed by
\citet{1980PollackCuzzi}.  The composition of the dust is 
that used in \citet{pollack_dust}, consisting of 
water, troilite, astronomical silicates, and organics.  
We adopt a size distribution for the dust of 
of $n(a) \propto a^{-3.5}$ where $a$ is the radius of the grain, 
with maximum and minimum radii of 
1 mm and 0.005 microns, respectively.
The large maximum grain size represents coagulation of grains 
in protoplanetary disks.  

\begin{figure}[htbp]
\plotone{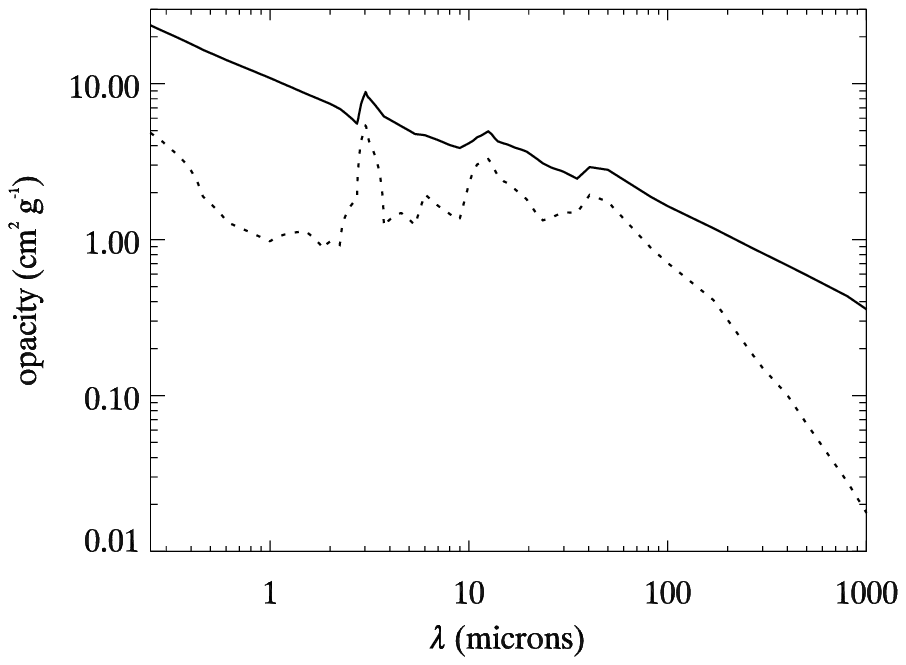}
\caption{\label{opacityplot}
Wavelength-dependent opacities for the dust model adopted for 
calculating simulated images.  
Dotted line: absorption only.  
Solid line: the total extinction (absorption+scattering).  
}
\end{figure}

In \figref{opacityplot}, we show the absorption 
(dotted line) and total extinction (solid line) 
coefficients in units of $\mbox{cm}^2\,\mbox{g}^{-1}$
versus wavelength for our adopted dust model.  
The total extinction includes both scattering and absorption.  
If $\kappa_{\nu}$ and $\chi_{\nu}$ 
are the frequency-dependent absorption and 
extinction coefficients, respectively, then the 
albedo is ratio of scattering to total extinction, so 
$\omega_{\nu} = 1 - \kappa_{\nu}/\chi_{\nu}$.
Since we assume that the dust is well-mixed, the 
dust density simply scales with the gas density.  

\subsection{Scattered Light Imaging}

We calculate scattered light from the disk assuming that it is face on 
and that stellar irradiation is primarily scattered from 
part of the disk that is optically thin to stellar irradiation.  
The optical depth to stellar light at a frequency $\nu$ is 
\begin{equation}
\tau_{\nu,*} = \int_{\ell} \chi_{\nu} \rho \, dl 
\end{equation}
where $\chi_{\nu}$ is the extinction coefficient, 
$\rho$ is the density of the disk, and $\ell$ is the path 
between the star and the point in the disk.  From this point 
on, the dependence on frequency, $\nu$, will be implicit and we will 
omit the subscripts.
We define the surface of the disk, $z_s(r)$ 
to be where $\tau_* = 2/3$. 
The cosine of the angle of incidence at radius $r$ is 
% \beta = z' * r/z
% z*\beta = z' * r
\begin{equation}
\mu(r) = \frac{( z_s' - z_s/r ) }{
\left[(z_s')^2 + 1\right]^{1/2}\left(1+z_s^2/r^2\right)^{1/2}} + \frac{4 R_*}{
3\pi \left(r^2+z_s^2\right)^{1/2}}. 
\end{equation}
The first term is a purely geometric term, 
and the second term accounts for the finite size of the star, in constrast 
to illuination by a point source.  
%%%%%% spline interpolation between log(z_s) and log(r) for calculating both z_s and slope
The slope of the surface is defined as $z_s' = d z_s/dr$.  
We use a log-log interpolation to calculate both 
$z_s$ and $z_s'$ as a function of $r$.  

Assuming isotropic scattering,  
the brightness of disk as a function of frequency, $I_{\nu}^s$, 
can be calculated by integrating the following equation 
from \citet{dalessio2}:
\begin{equation}\label{fscatDAlessio}
\frac{dI^{s}}{dZ} = 
-\frac{(\chi-\kappa) R_*^2}{4l^2} B(T_*) \exp[-\tau_*
-\tau_{\mbox{\scriptsize{obs}}}(Z)]
\end{equation}
where $l$ is the distance to the star, 
%which we infer this to be equal to $(\chi_{\nu}-\kappa_{n\u})\rho$;
$B(T_*)$ is the Planck function at the effective temperature of the star,
and $\tau_{\mbox{\scriptsize{obs}}}$ is the optical depth 
to the observer, 
\(
\tau_{\mbox{\scriptsize{obs}}} = 
\int_{\mbox{\scriptsize{obs}}} \chi_{\nu} \rho \, dl, 
\)
integrated along the line-of-sight from the observer to the disk.  
For a face-on disk, this is integrated 
along the vertical axis, so $Z=z$.  

At low optical depths, $\rho$ is small, so contributions to 
$I^{s}$ are small.  At high optical 
depths, the exponential term goes to $0$, so those contributions 
are also small.  Only the regions close to the surface contribute 
significantly to the scattered light intensity, so we assume 
$l \approx \sqrt{r^2+z_s^2}$.  
We assume that the disk is locally plane parallel at the surface, 
so that if $i$ is the angle between the line of sight to 
the observer and normal to the surface, 
$\tau_* = \cos i \, \tau_{\mbox{\scriptsize{obs}}}(z)/\mu$. 
Note that $i$ is not the inclination of the disk, but 
rather the angle with respect to the disk surface, which is 
not generally parallel to the ecliptic.  
Then Eq.~(\ref{fscatDAlessio}) becomes 
\begin{equation}
\frac{1}{\chi\rho}\frac{dI^{s}}{dz} =
\frac{dI^{s}}{d\tau} =
\frac{\omega R_*^2}{4(r^2+z_s^2)} B(T_*) \exp[-(1+\cos i/\mu)\tau].
\end{equation}
Integrating from $\tau = 0\rightarrow\infty$, 
\begin{equation}\label{iscatt}
I^{\mbox{\scriptsize{scatt}}} = 
\frac{\omega \mu R_*^2 B(T_*) }{4(r^2+z_s^2)(\mu+\cos i)}.
\end{equation}
%% The observed brightness of a star at distance $d$ is 
%% \begin{equation}
%% F_{\mbox{\scriptsize{obs}}} = \pi B \left(\frac{R_*}{d}\right)^2
%% \end{equation}
%% so we can write express the surface brightness in scattered light 
%% in units of the apparent brightness of the star per square arcsecond:
%% \begin{equation}\label{scattered}
%% I^{\mbox{\scriptsize{scatt}}} = 
%% \frac{\omega_\nu \mu }{4\pi(\mu+\iota)}
%% \left(\frac{d}{\mbox{pc}}\right)^2
%% \left(\frac{r^2+z_s^2}{\mbox{AU}^2}\right)^{-1}
%% \left(\frac{F_{\mbox{\scriptsize{obs}}}}{\mbox{asec}^2}\right),
%% \end{equation}
%% which has units of Jy/asec$^2$.

\subsection{Thermal Emission}

At longer wavelengths, thermal emission from the disk material become 
important.  
The thermal emission from a disk is calculated following 
\citet{dalessio2}, by integrating the equation
\begin{equation}
\frac{dI^t}{dZ} = \kappa \rho B(T_d) \exp(-\tau\sub{obs})  
\end{equation}
along the line of sight, 
where $T_d$ is the local disk temperature and 
$B(T_d)$ is the thermal emission at that temperature.  
The thermal emission and scattered light are summed to 
give the total disk surface brightness:
\begin{equation}
I^{\mbox{\scriptsize{disk}}} = I^s + I^t.
\end{equation}

%% \subsection{Anisotropy and Polarization}

%% A common method of modeling anisotropic scattering is by adopting 
%% the Henyey-Greenstein phase function \citep{1941HenyeyGreenstein}:
%% \begin{equation}
%% f(g,\theta) = \frac{1-g^2}{4\pi\left( 1 + g^2 - 2g\cos\theta \right)^{3/2}}
%% \end{equation}
%% where $\theta$ is the scattering angle; and 
%% $g$ is asymmetry parameter, which can range from $-1$ to $1$, negative 
%% values for back scattering and positive values for forward scattering.
%% This equation replaces the factor of $1/(4\pi)$ in 
%% Eq.~(\ref{scattered}), for the case of $g=0$ for isotropic scattering.  

%% The fractional linear polariazation from Rayleigh scattering is 
%% \citep{1974HansenTravis} 
%% \begin{eqnarray}
%% \frac{I\sub{pol}}{I} &=& \frac{(Q^2 + U^2)^{1/2}}{I} \\
%% &=& 
%% \frac{\sin^2 \theta}{1 + \cos^2 \theta}.
%% \end{eqnarray}
%% We assume that circular polarization is negligible, so $V=0$.  
%% Here, $I$, $Q$, $U$, and $V$ are the canonical Stokes parameters.  

\section{Results}\label{results}

\begin{figure}
\plottwo{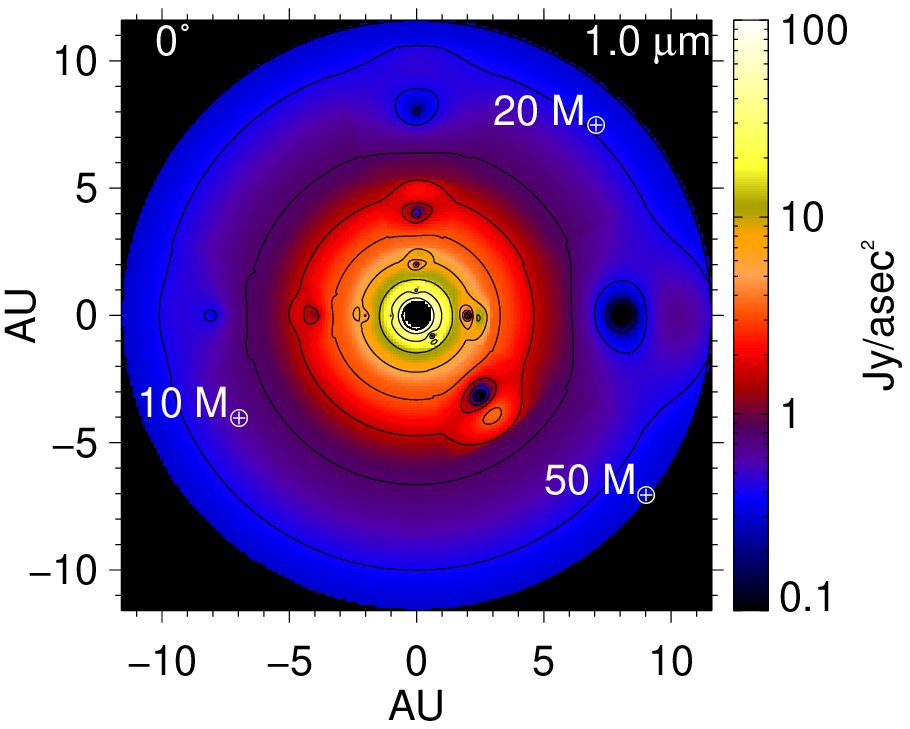}{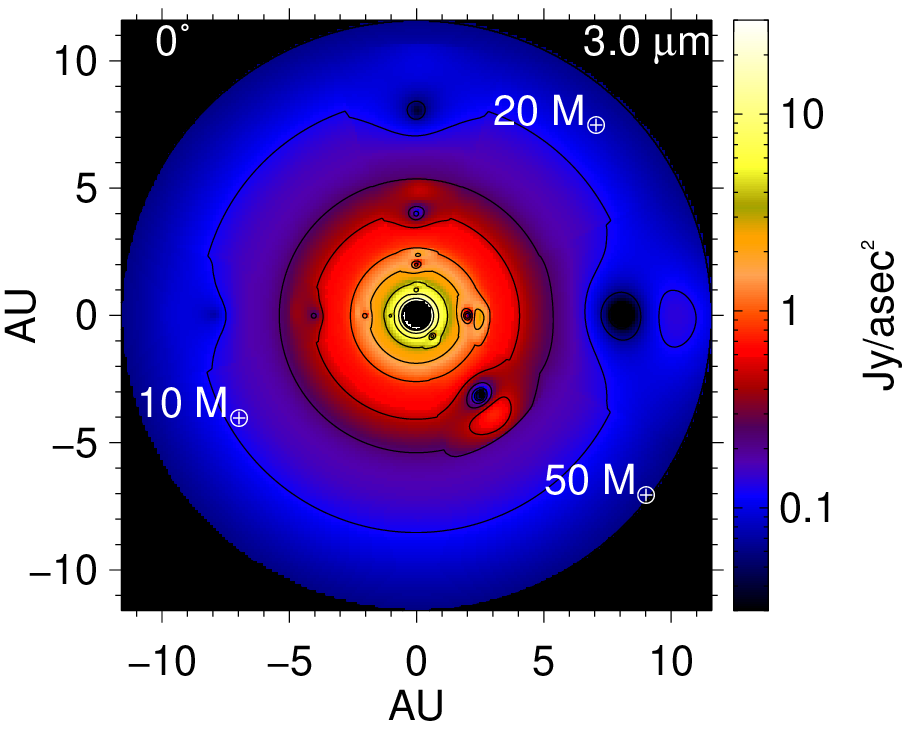}\\
\plottwo{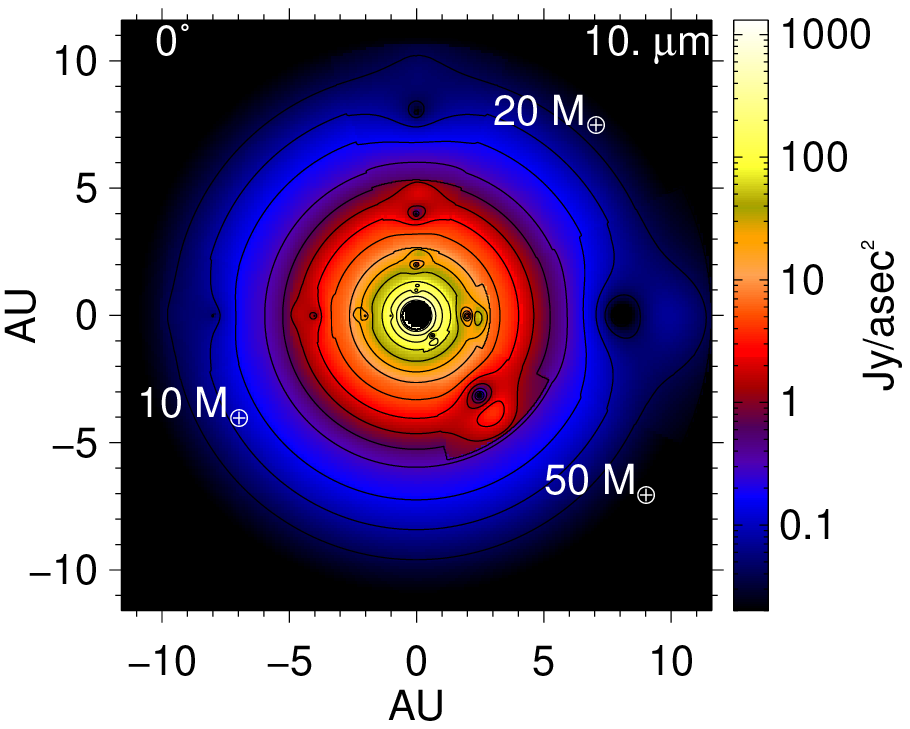}{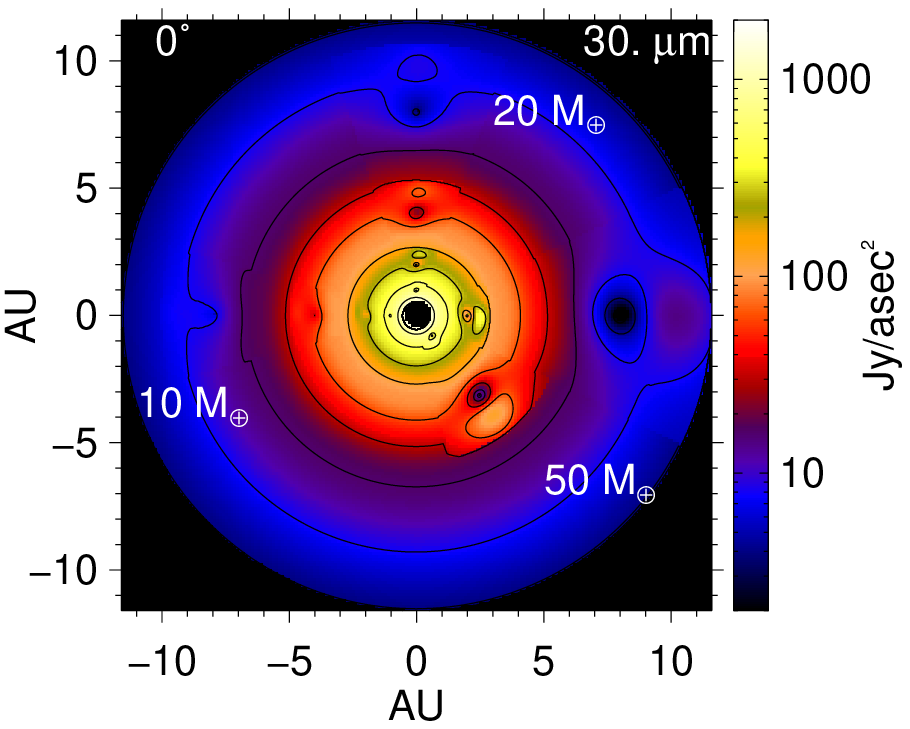}\\
\plottwo{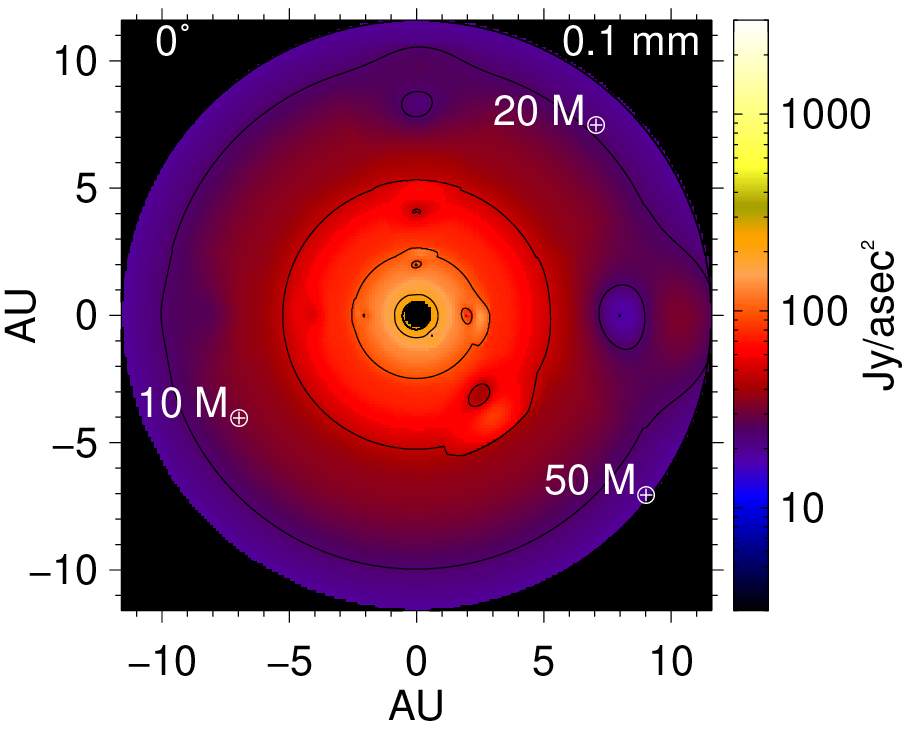}{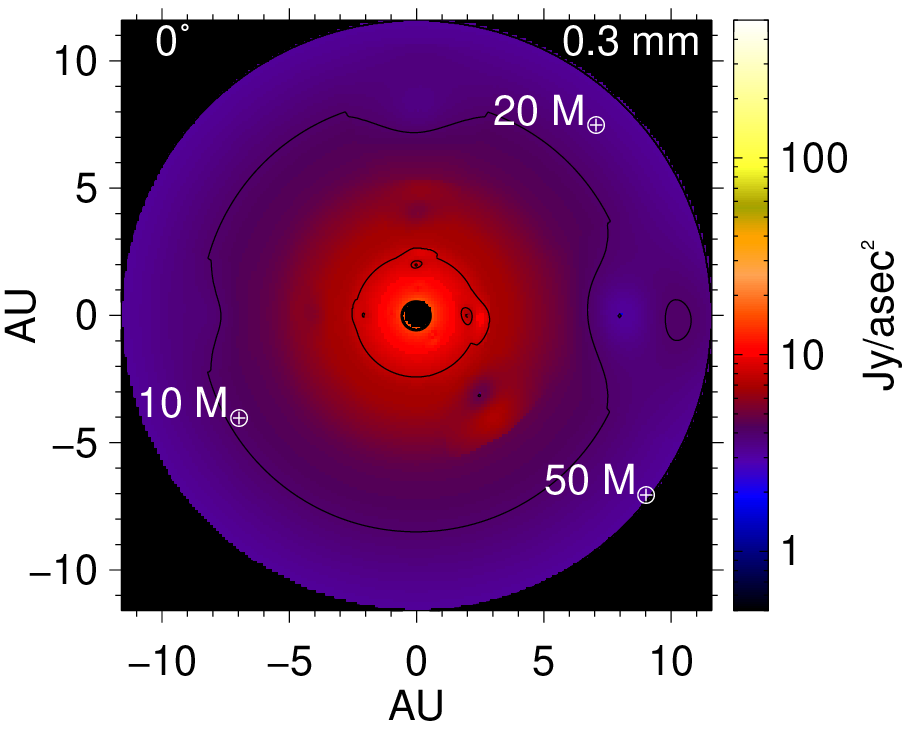}
\caption{\label{allfaceon}
Planets of varying masses embedded in a face-on disk at 
1, 2, 4 and 8 AU seen at varying wavelengths.  
From left to right, top to bottom: 
1, 3, 10, 30, 100, and 300 microns.  
Planets at the 9 o'clock position are 10 $M_{\earth}$, 
planets at the 12 o'clock position are 20 $M_{\earth}$, 
and planets at the 3-5 o'clock positions are 50 $M_{\earth}$.
}\end{figure}

\begin{figure}
\plotone{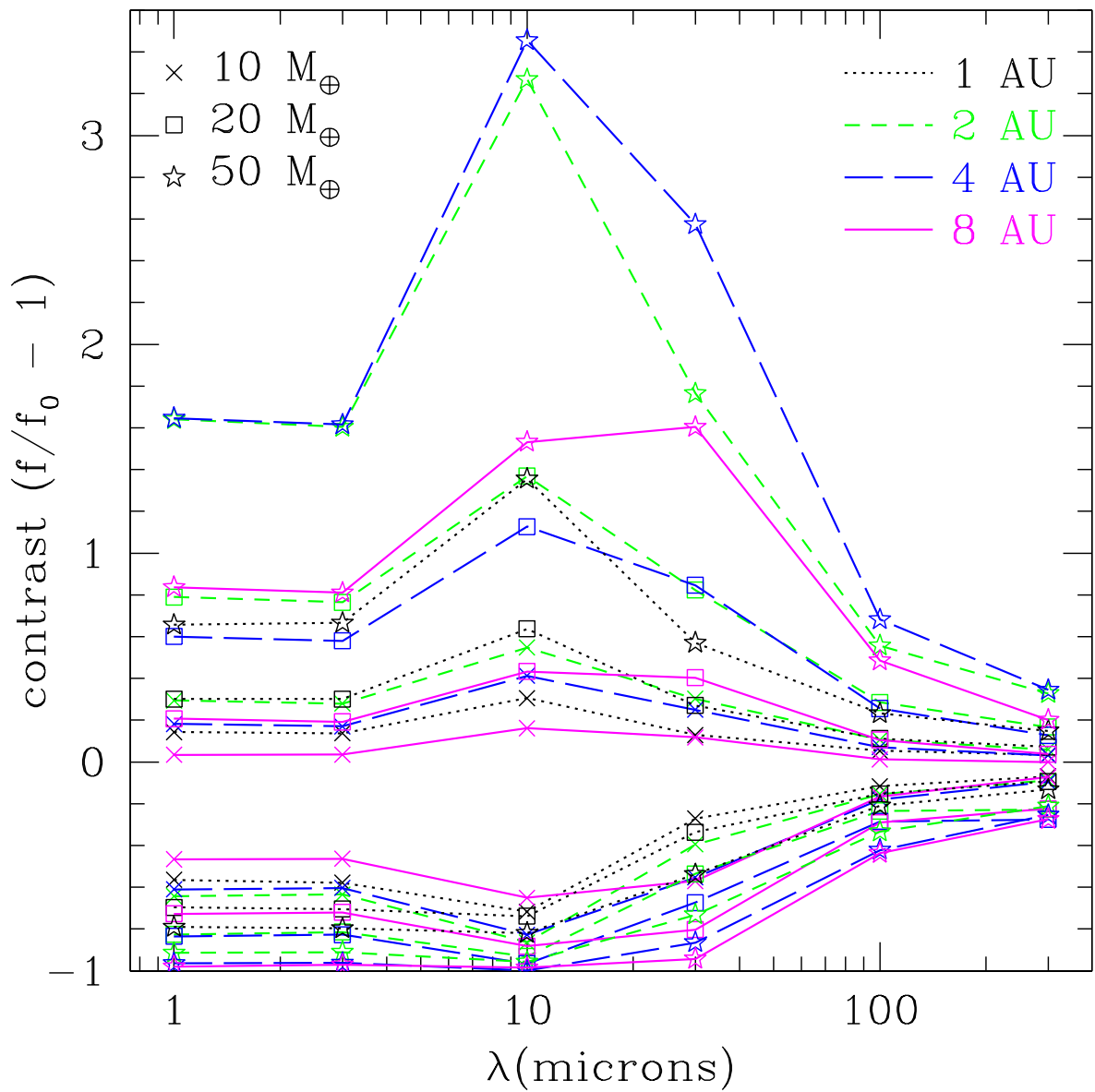}
\caption{\label{contrast} 
Minimum/Maximum contrast versus wavelength.
Orbital distances of 1, 2, 4, and 8 AU are represented by 
black dotted, green short-dashed, blue long-dashed, and 
solid magenta lines, respectively.
Planet masses are indicated by crosses/squares/stars for 
10/20/50 $M_{\earth}$.  
}
\end{figure}

\begin{figure}
\plotone{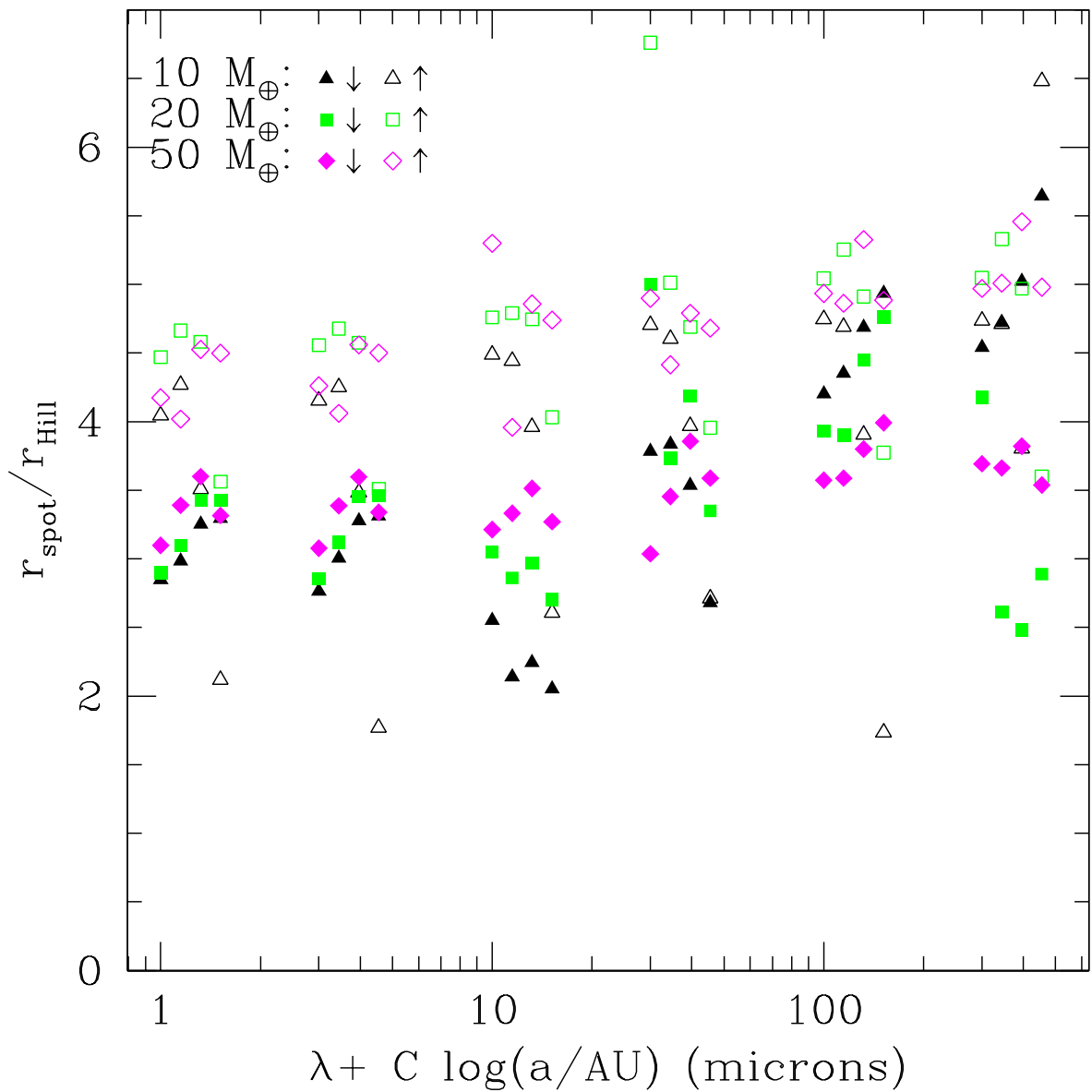}
\caption{\label{spotradii} 
Equivalent radii $r\sub{spot}$ of shadowed/brightened regions, 
scaled to the Hill radius $r\sub{Hill}$, versus wavelength.  
The points have been horizontally shifted according to 
to distinguish between orbital distances, but the wavlengths 
correspond to 1, 3, 10, 30, 100, and 300 $\mu$m exactly.  
Black triangles/green squares/magenta diamonds 
indicate planet masses of 
10/20/50 $M_{\earth}$.  Filled symbols indicate shadows and 
open symbols indicate bright spots. 
}
\end{figure}

\begin{figure}
\includegraphics[width=2.12in]{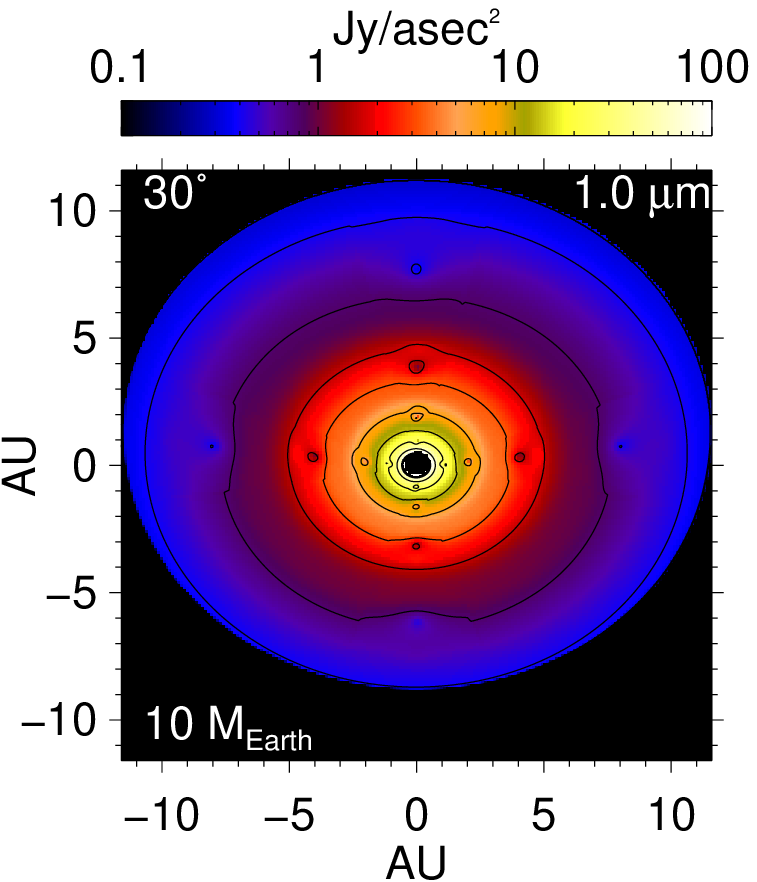}
\includegraphics[width=2.12in]{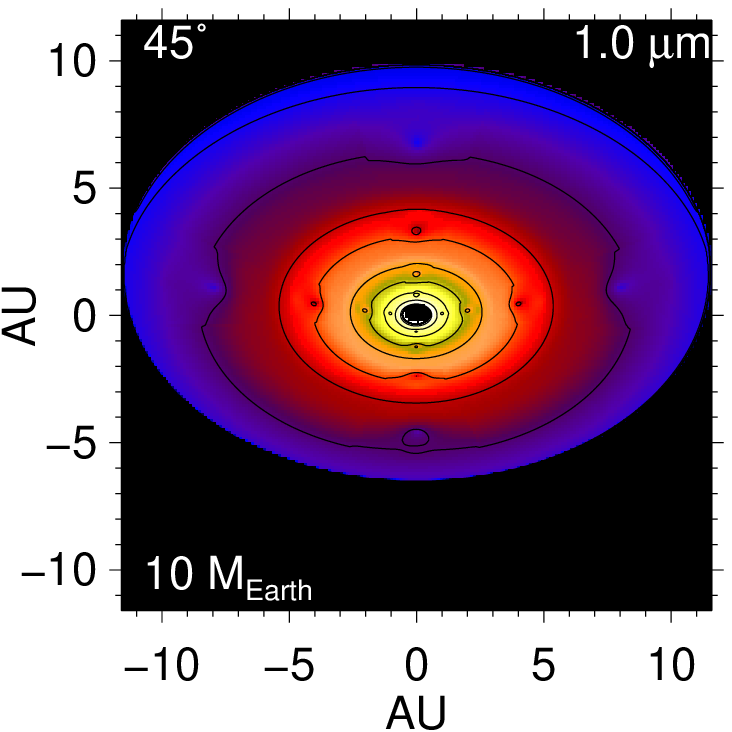}
\includegraphics[width=2.12in]{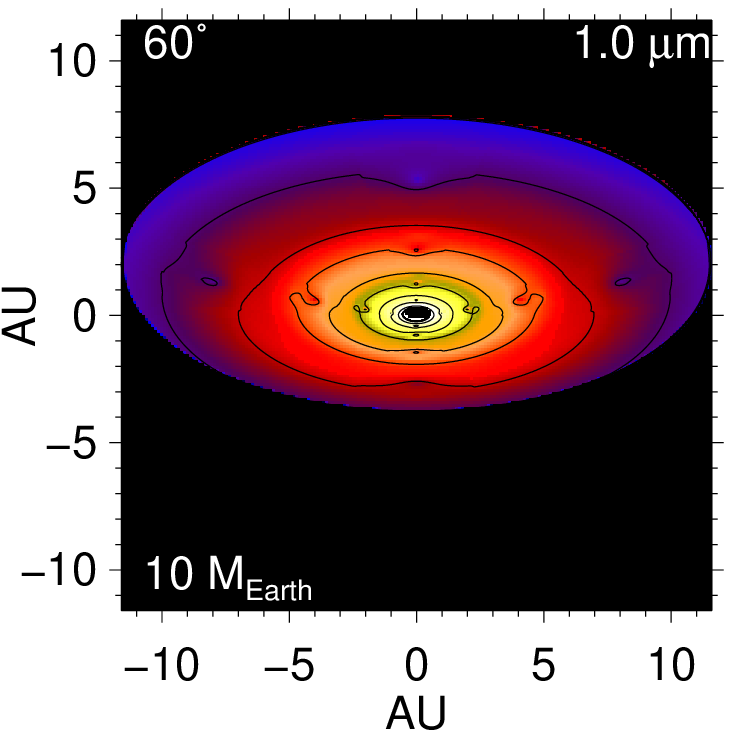}\\
\includegraphics[width=2.12in]{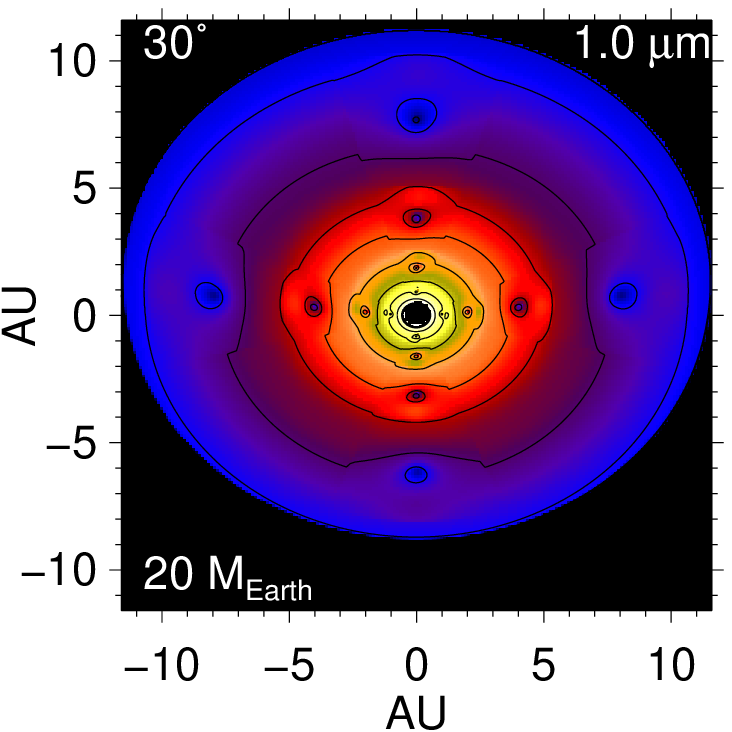}
\includegraphics[width=2.12in]{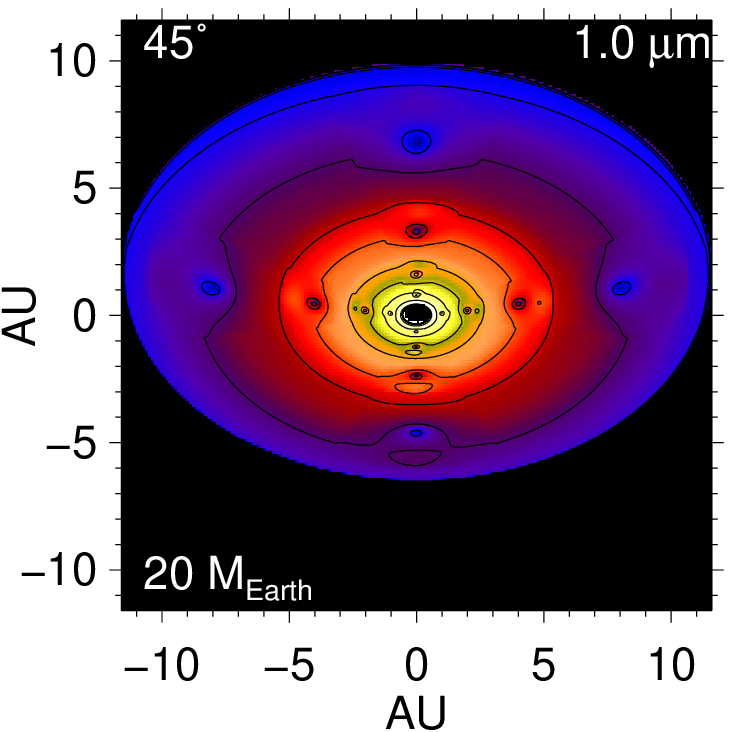}
\includegraphics[width=2.12in]{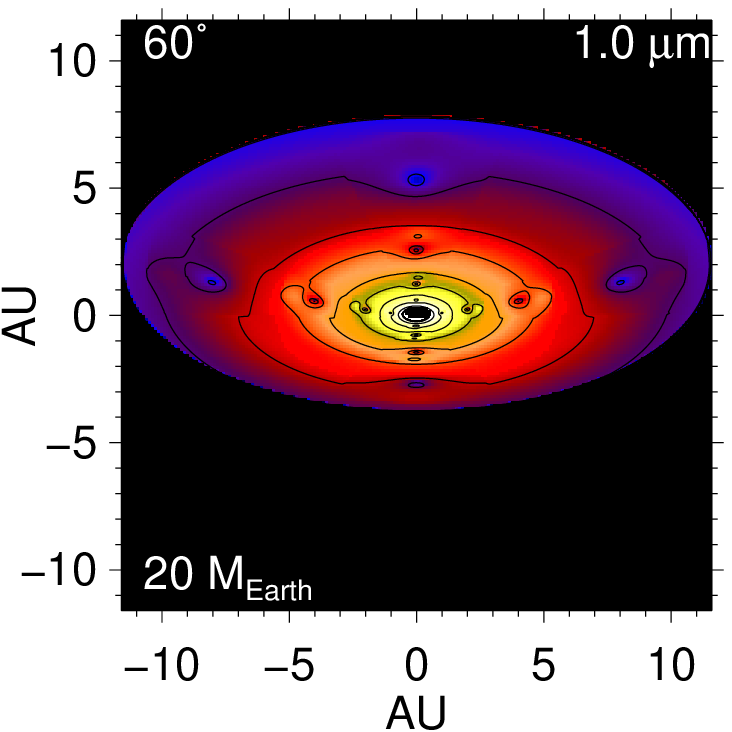}\\
\includegraphics[width=2.12in]{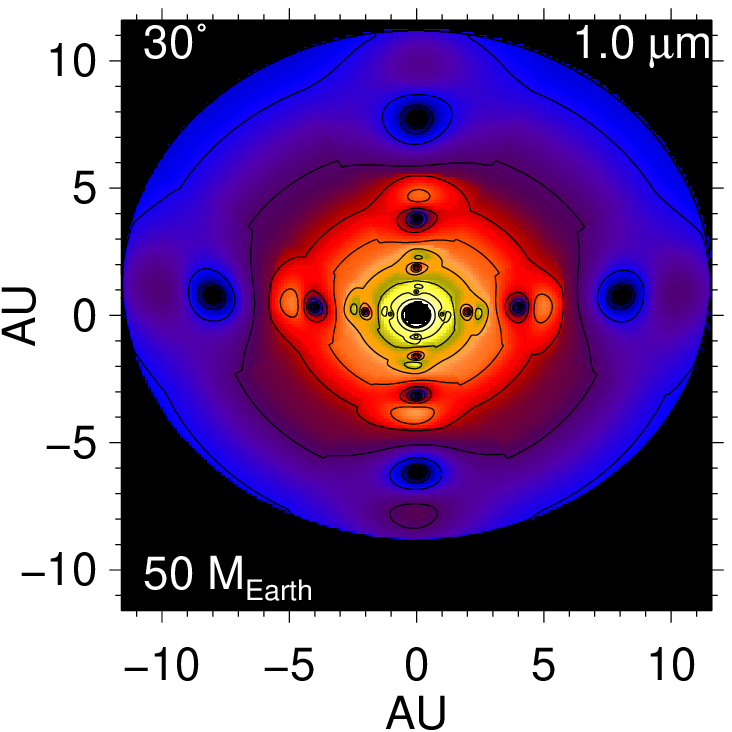}
\includegraphics[width=2.12in]{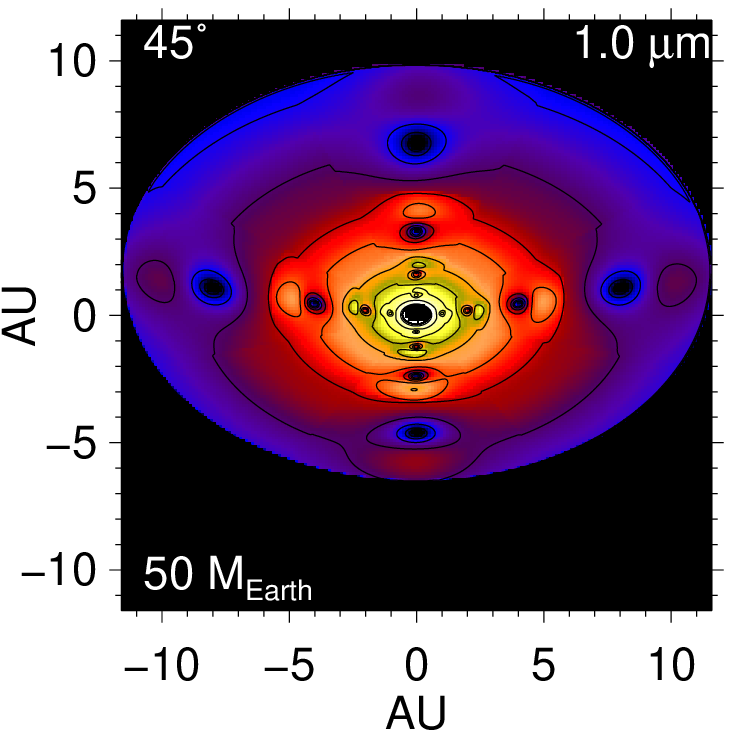}
\includegraphics[width=2.12in]{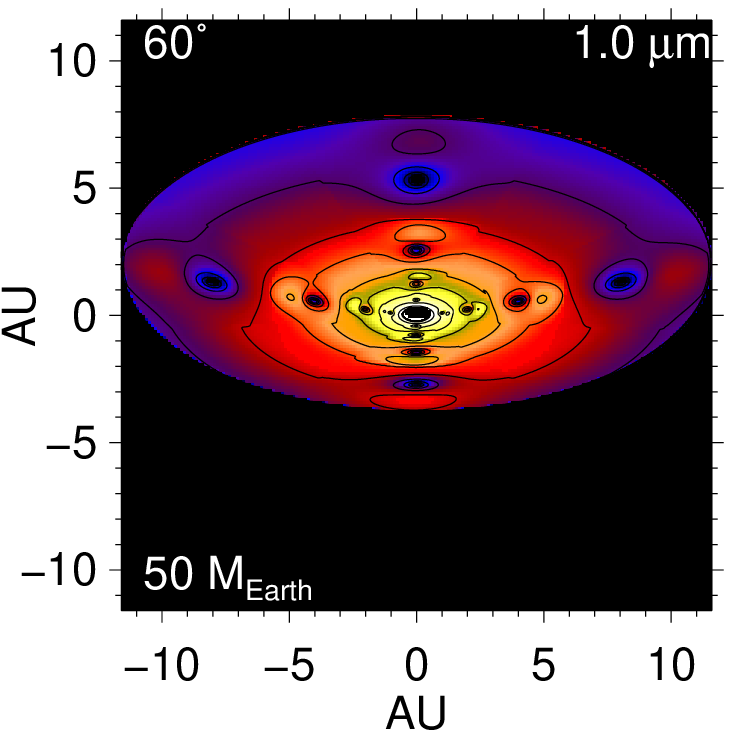}
\caption{\label{onemicronincl}
Simulated images of inclined disks perturbed by embedded planets at 
1 $\mu$m.  
The stellar brightness at 1 $\mu$m is 1.54 Jy at a distance of 100 pc. 
Each image shows how a planet of a given mass will perturb the appearance 
of the disk at 1, 2, 4, and 8 AU, and at phases
0, $\pi/2$, $\pi$, and $3\pi.2$.  
From left to right, the disks are inclined at 30, 45, and 60$\degr$.
The planet masses from top to bottom are 10, 20 and 50 $M_{\earth}$.  
}
\end{figure}

\begin{figure}
\includegraphics[width=2.12in]{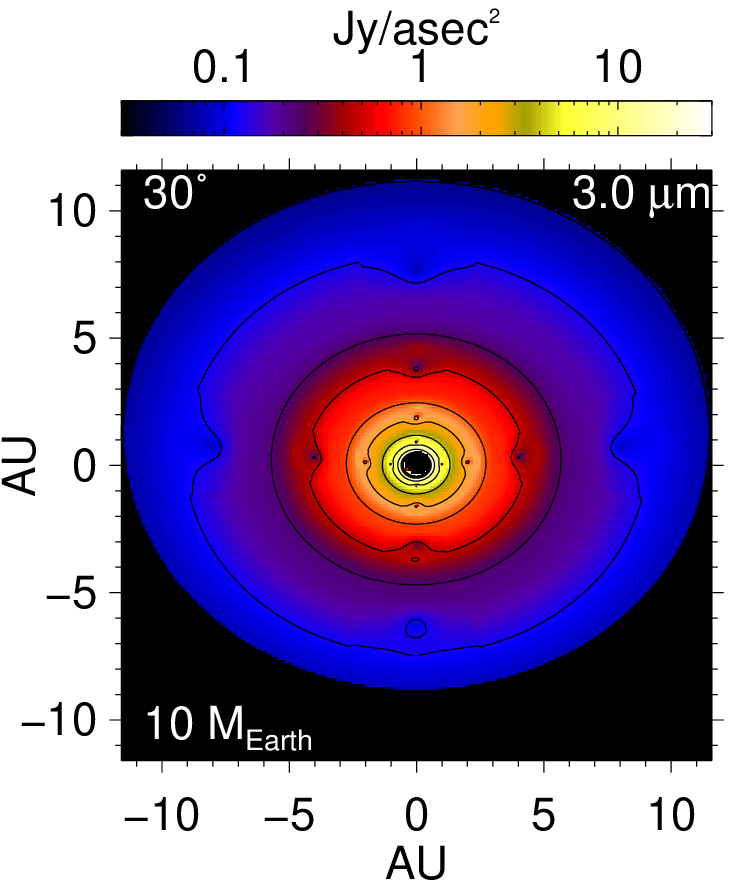}
\includegraphics[width=2.12in]{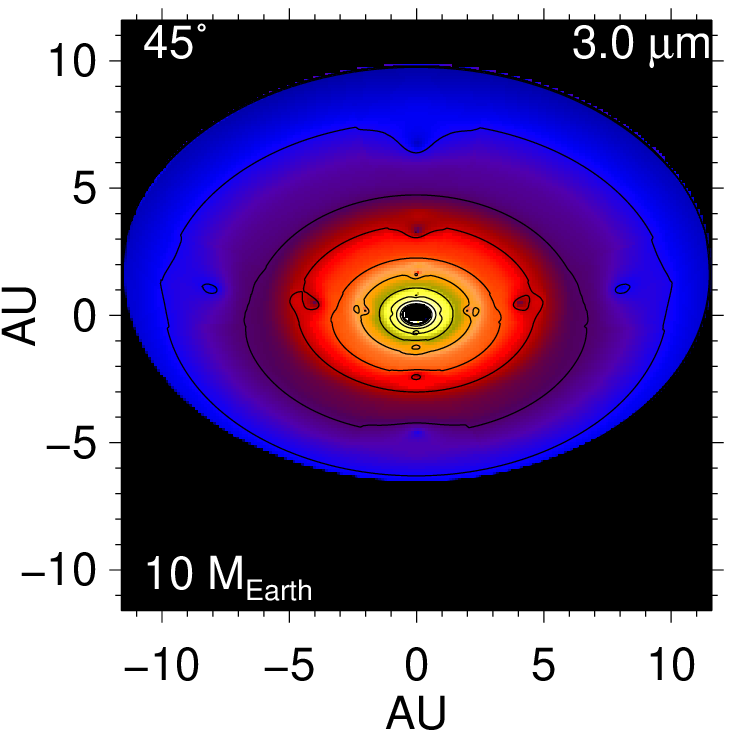}
\includegraphics[width=2.12in]{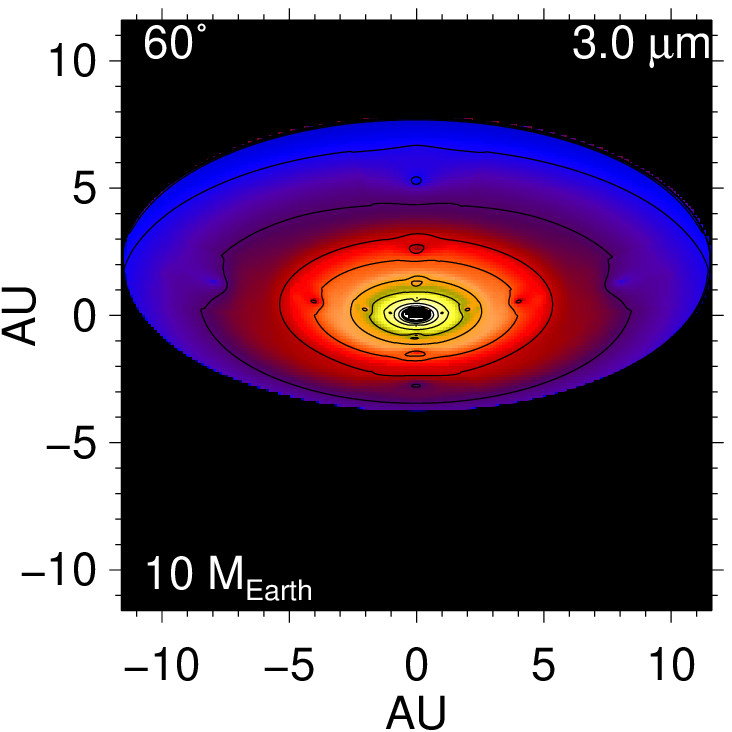}\\
\includegraphics[width=2.12in]{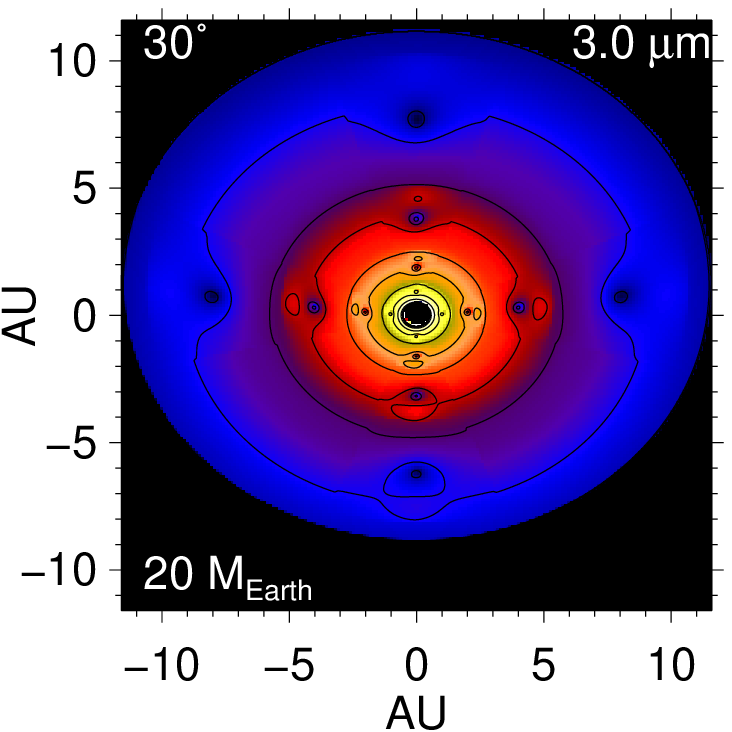}
\includegraphics[width=2.12in]{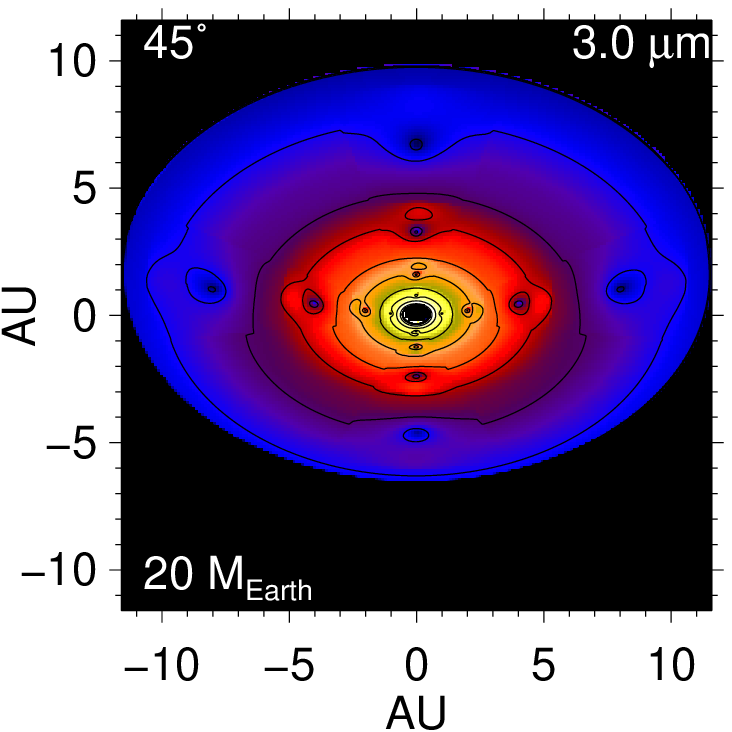}
\includegraphics[width=2.12in]{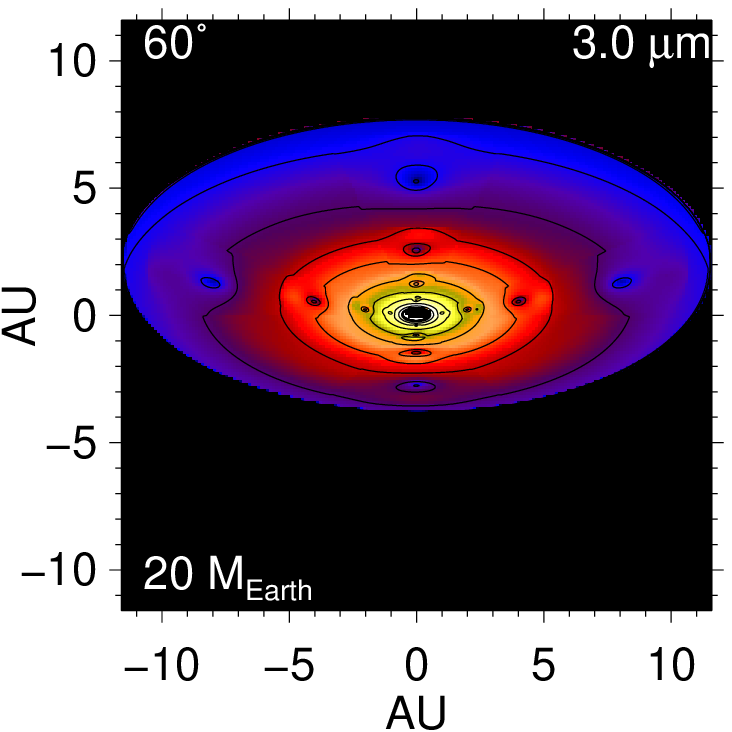}\\
\includegraphics[width=2.12in]{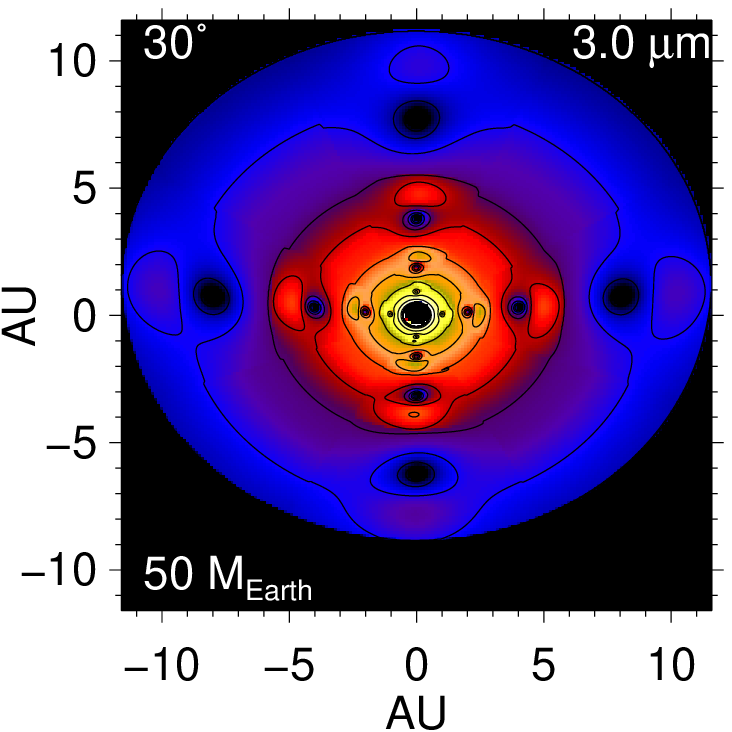}
\includegraphics[width=2.12in]{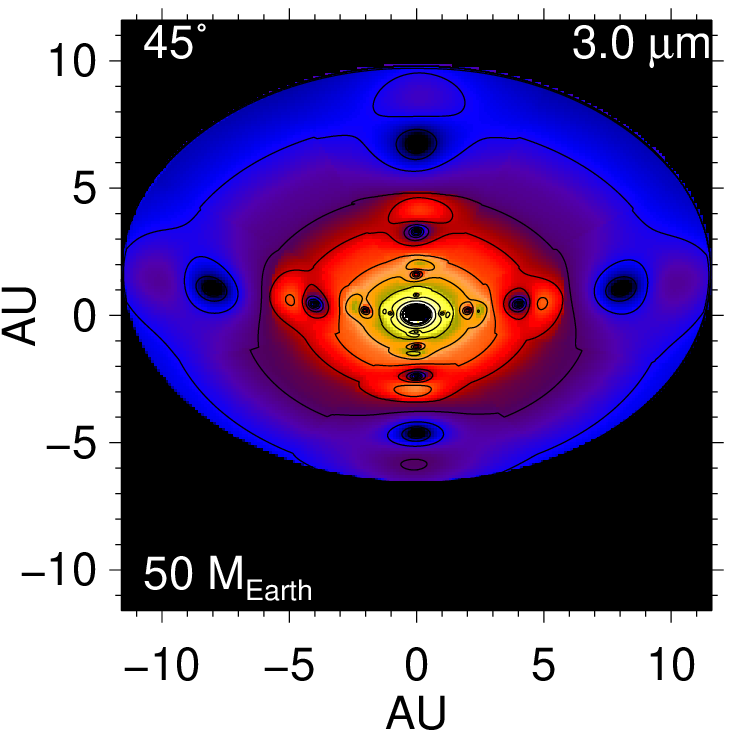}
\includegraphics[width=2.12in]{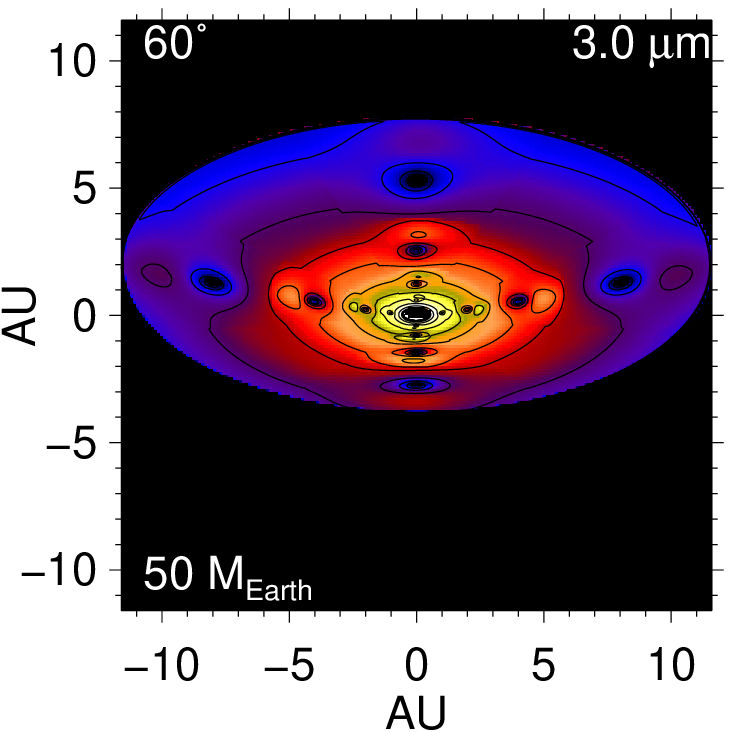}
\caption{\label{threemicronincl}
Simulated images of inclined disks perturbed by embedded planets at 
3 $\mu$m.  
The stellar brightness at 3 $\mu$m is 0.77 Jy at a distance of 100 pc. 
See \figref{onemicronincl} for description.  
}
\end{figure}

\begin{figure}
\includegraphics[width=2.12in]{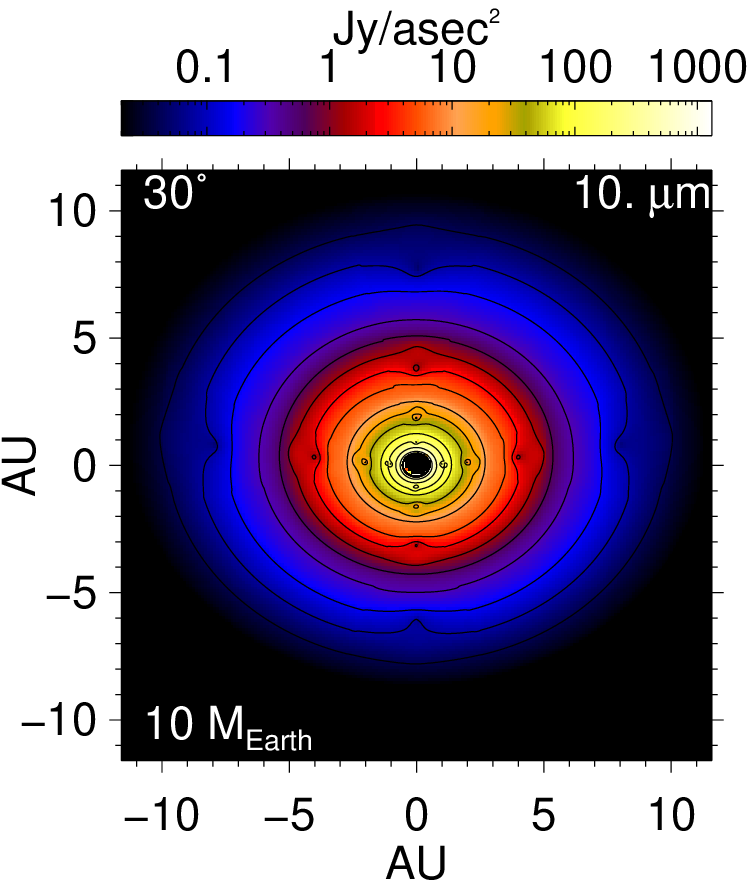}
\includegraphics[width=2.12in]{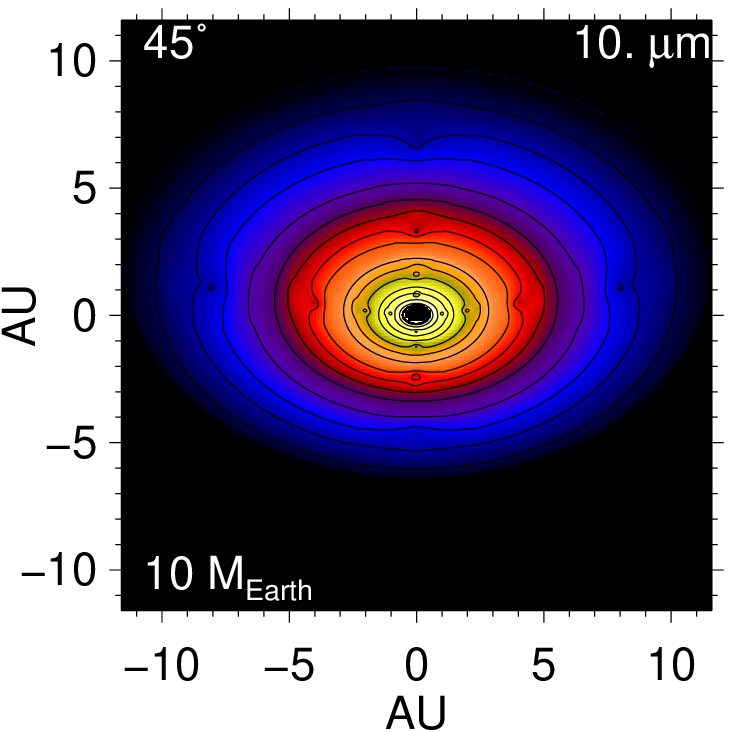}
\includegraphics[width=2.12in]{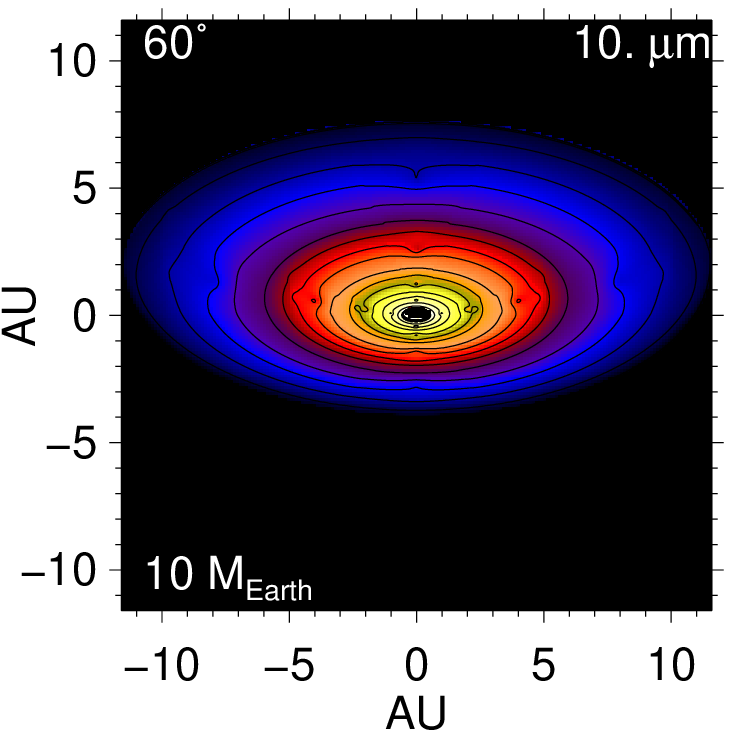}\\
\includegraphics[width=2.12in]{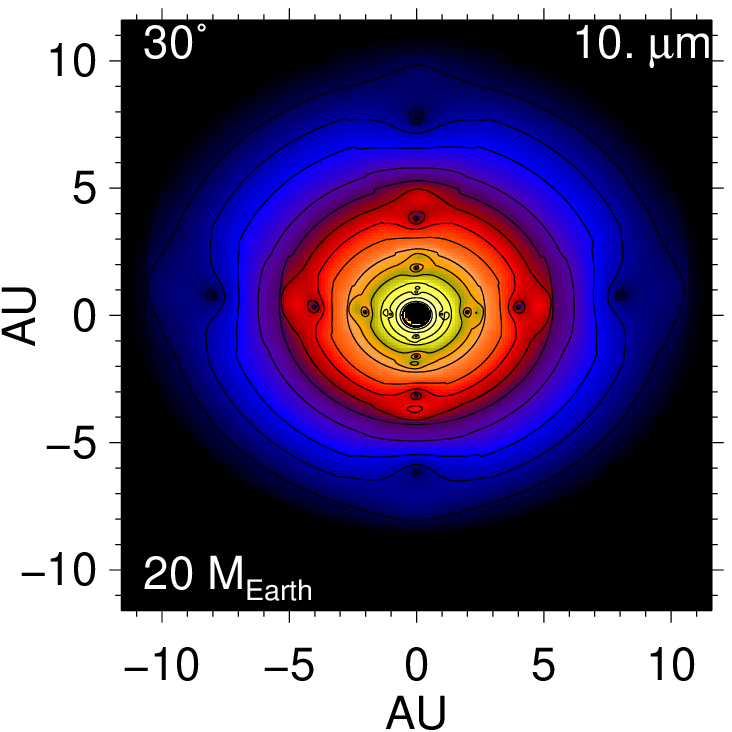}
\includegraphics[width=2.12in]{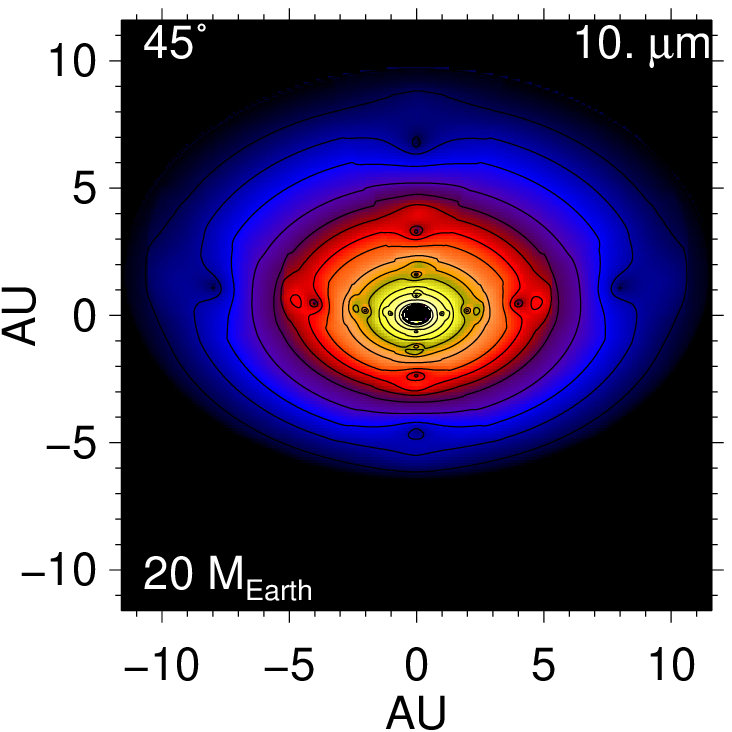}
\includegraphics[width=2.12in]{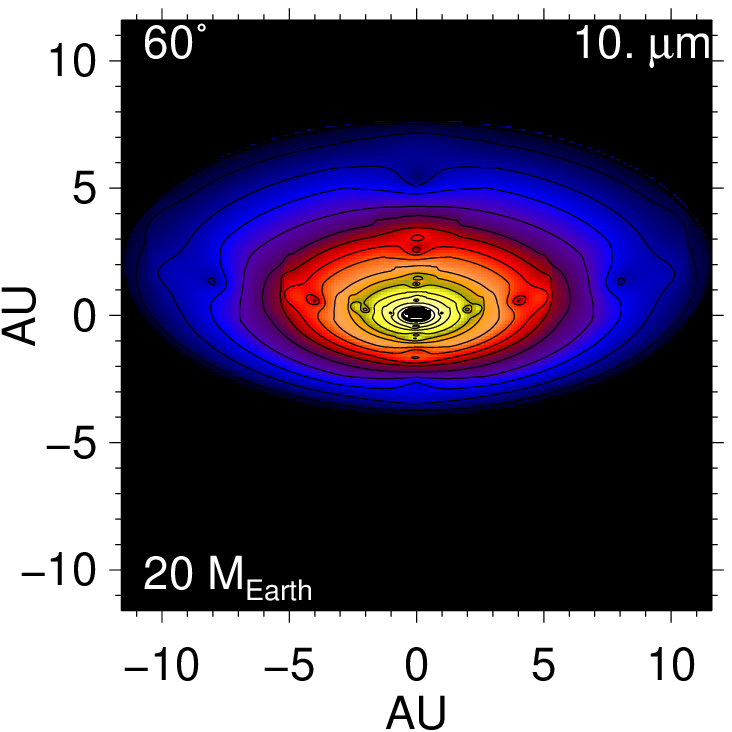}\\
\includegraphics[width=2.12in]{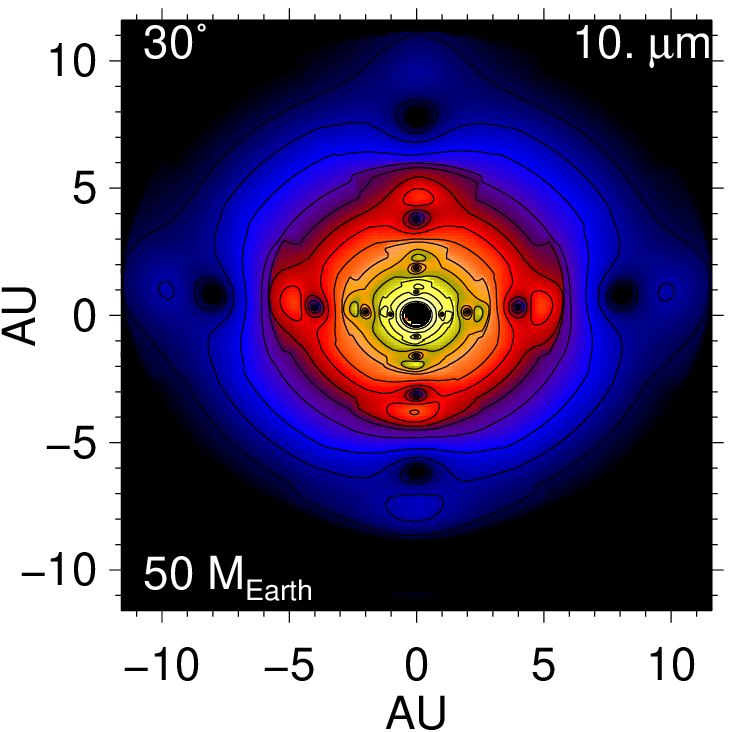}
\includegraphics[width=2.12in]{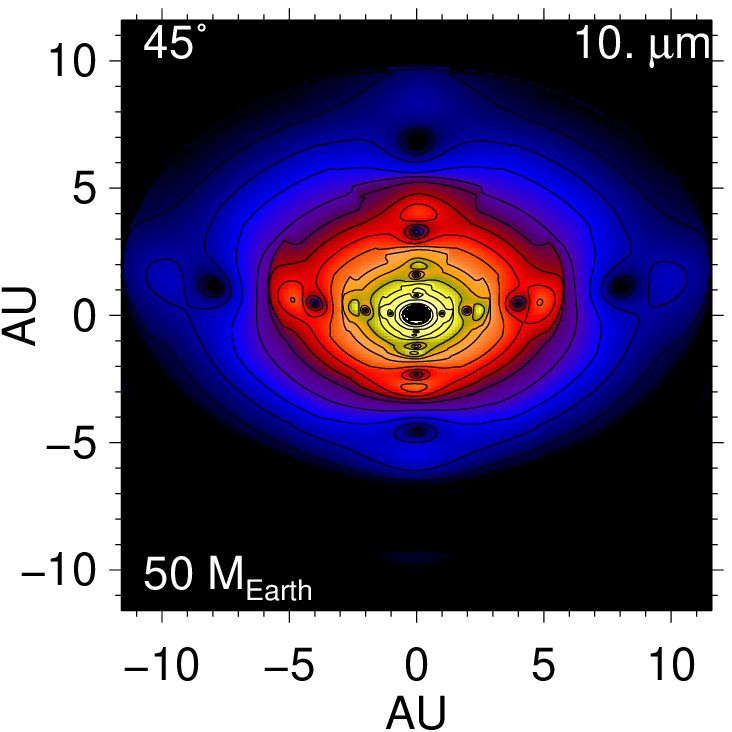}
\includegraphics[width=2.12in]{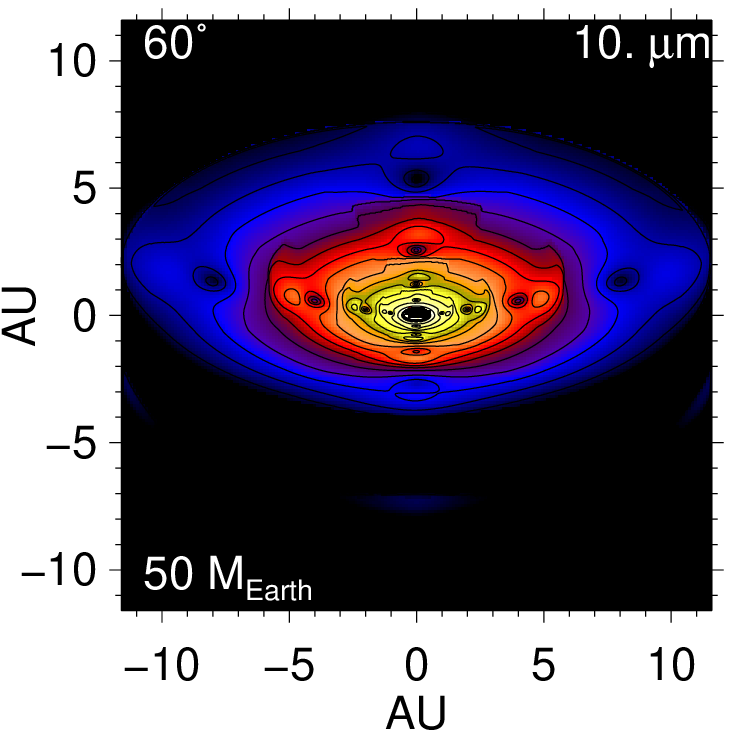}
\caption{\label{tenmicronincl}
Simulated images of inclined disks perturbed by embedded planets at 
10 $\mu$m.  
The stellar brightness at 10 $\mu$m is 0.107 Jy at a distance of 100 pc. 
See \figref{onemicronincl} for description.  
}
\end{figure}

\begin{figure}
\includegraphics[width=2.12in]{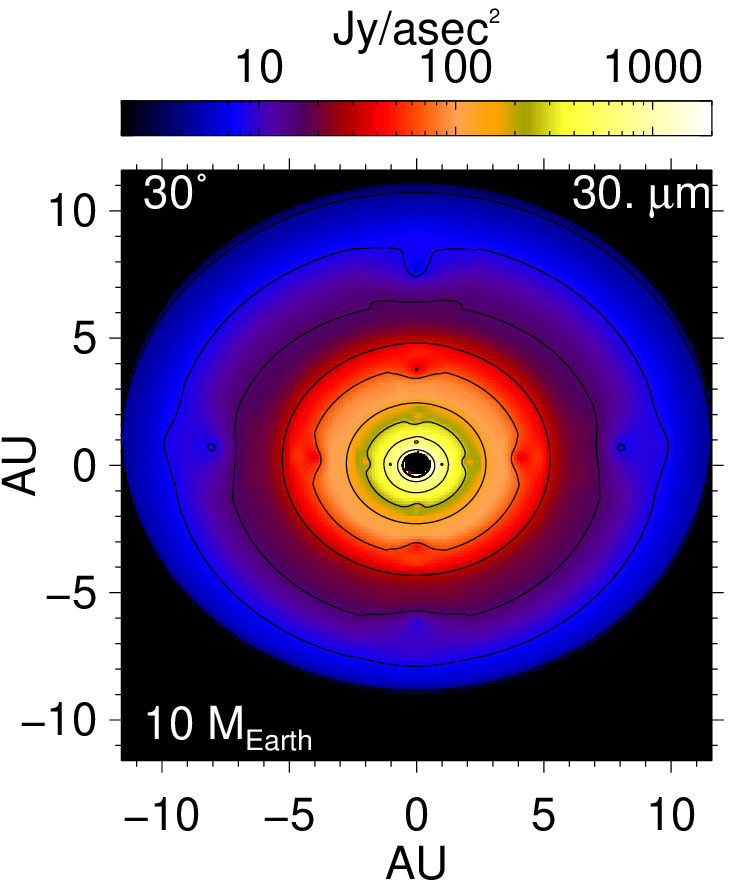}
\includegraphics[width=2.12in]{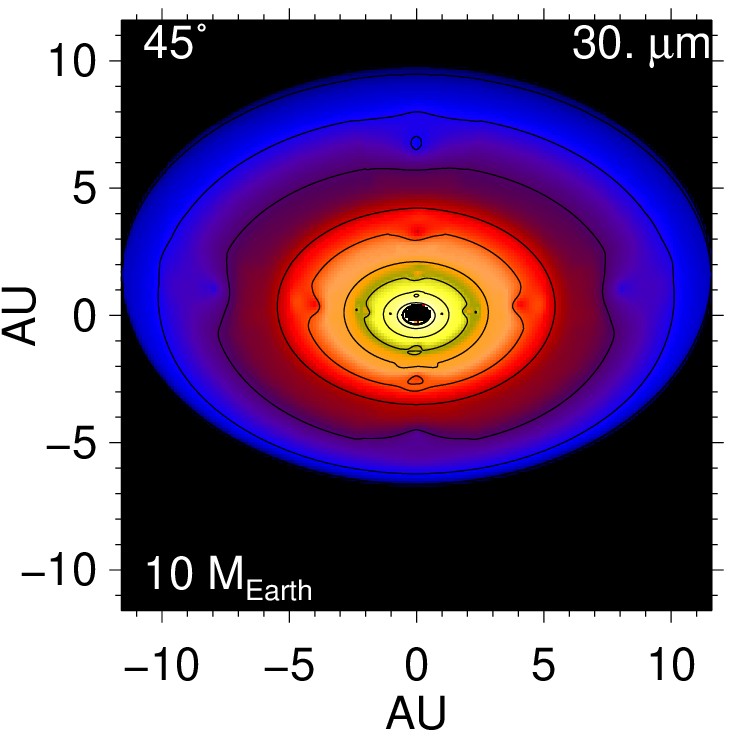}
\includegraphics[width=2.12in]{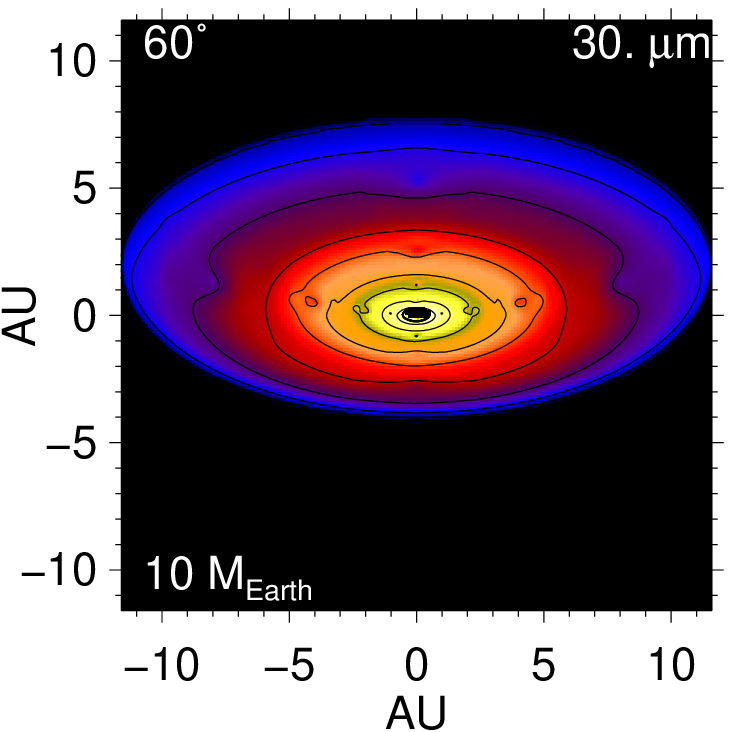}\\
\includegraphics[width=2.12in]{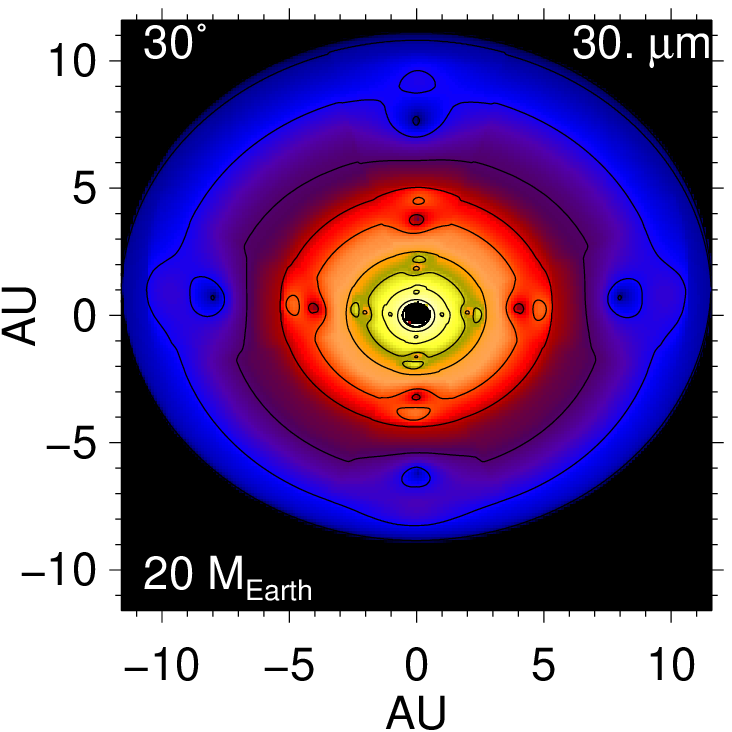}
\includegraphics[width=2.12in]{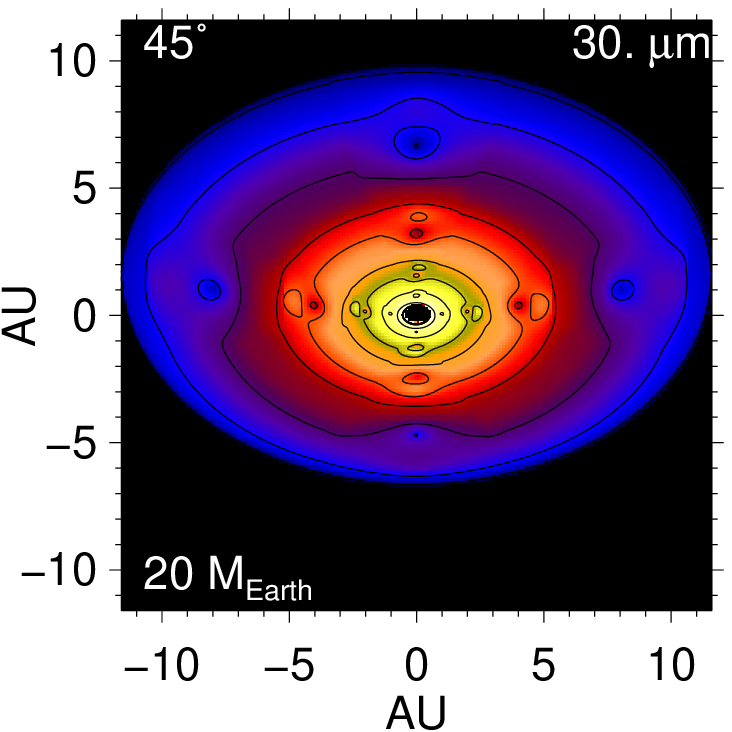}
\includegraphics[width=2.12in]{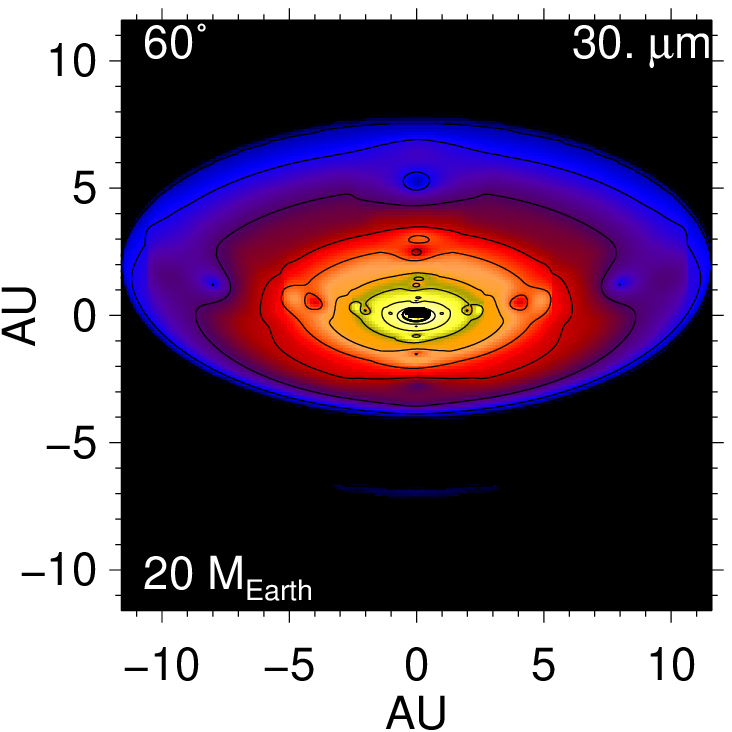}\\
\includegraphics[width=2.12in]{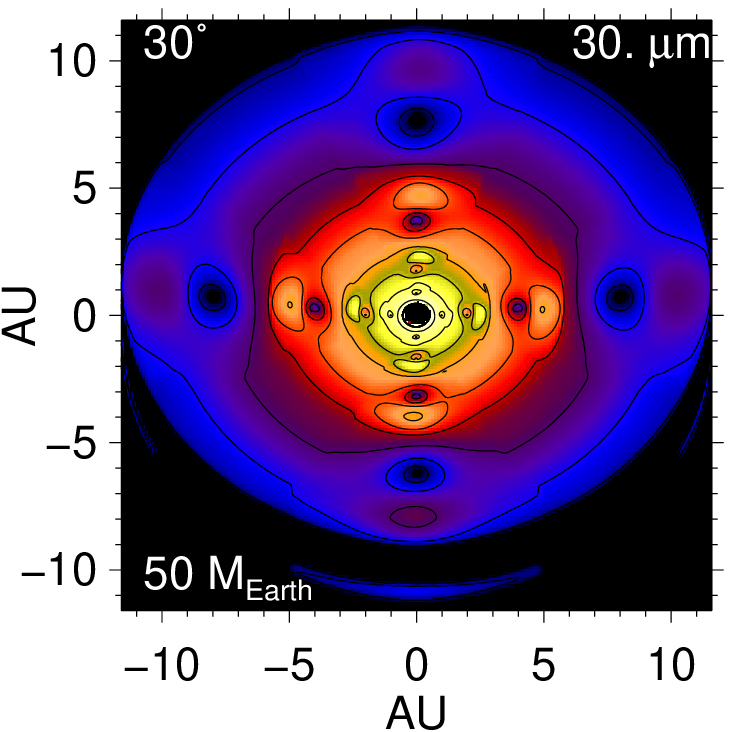}
\includegraphics[width=2.12in]{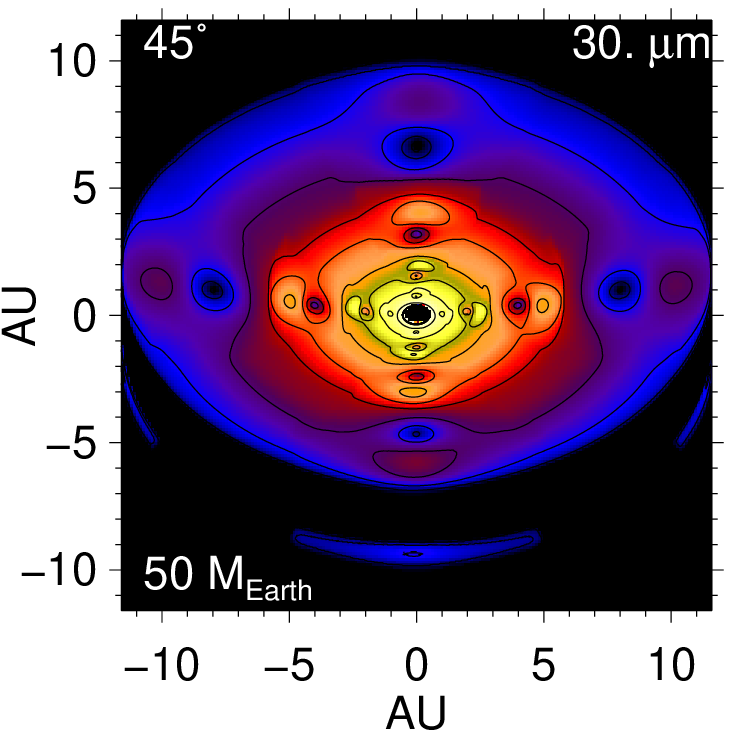}
\includegraphics[width=2.12in]{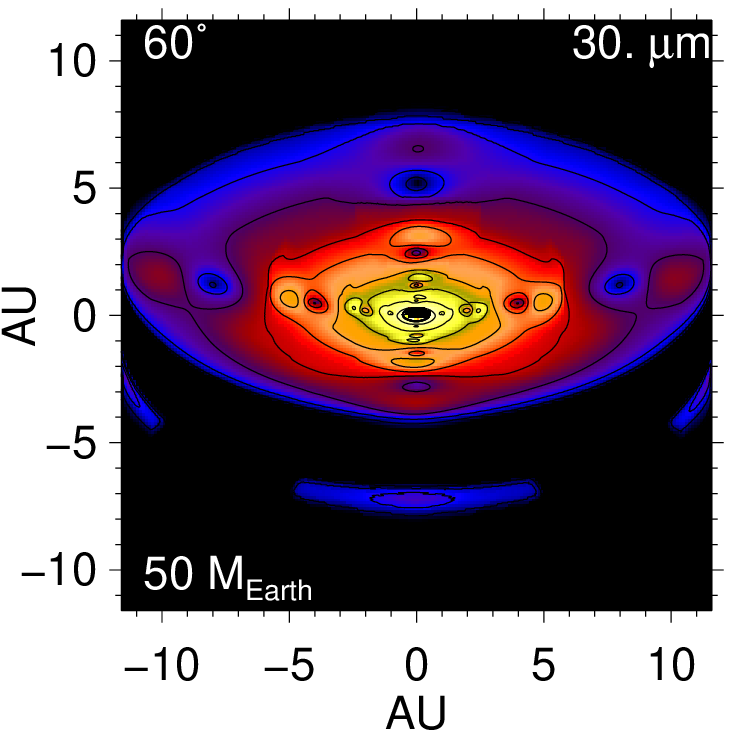}
\caption{\label{thirtymicronincl}
Simulated images of inclined disks perturbed by embedded planets at 
30 $\mu$m.  See \figref{onemicronincl} for description.  
}
\end{figure}

\begin{figure}
\includegraphics[width=2.12in]{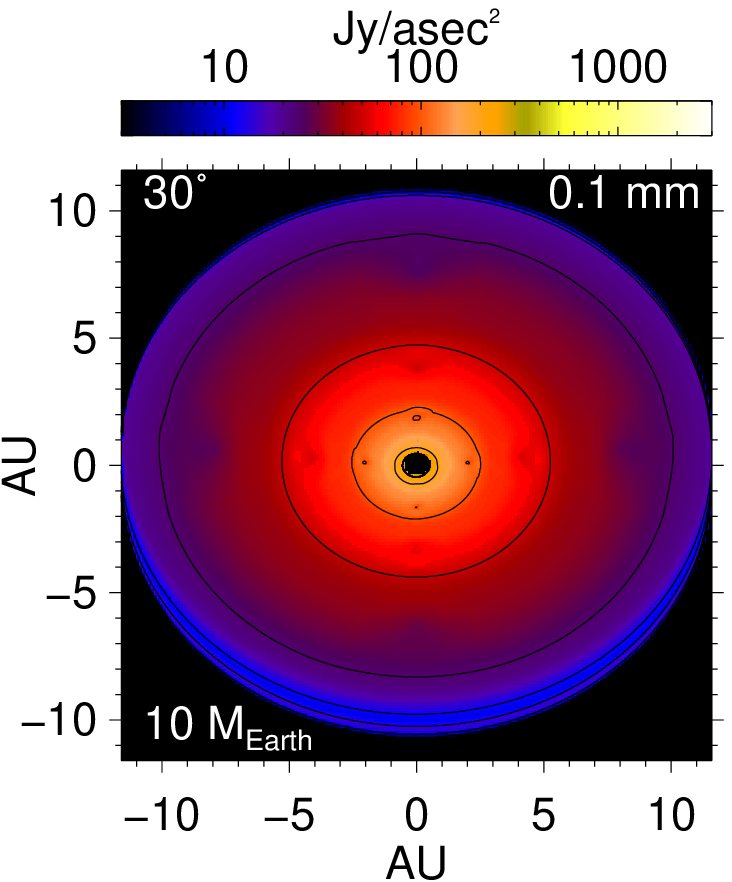}
\includegraphics[width=2.12in]{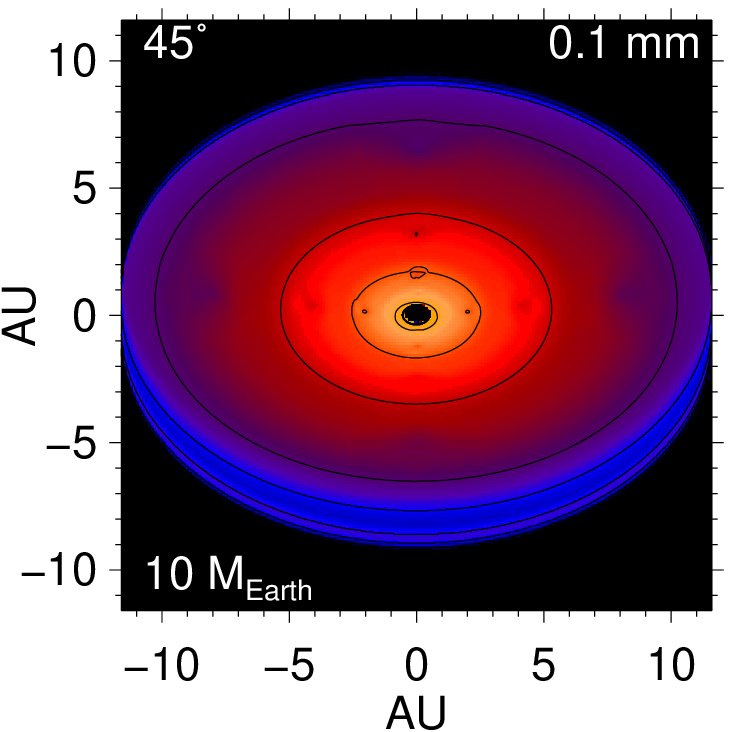}
\includegraphics[width=2.12in]{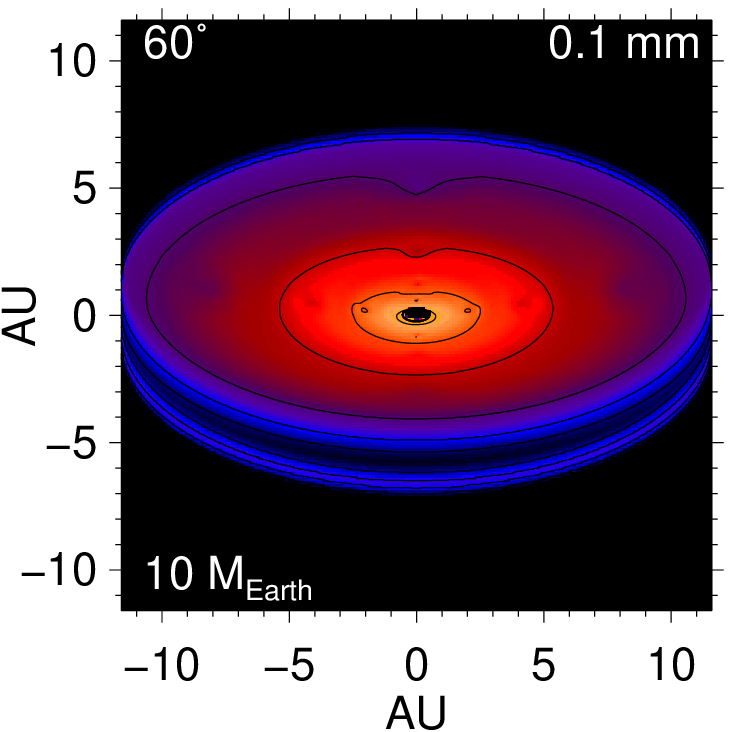}\\
\includegraphics[width=2.12in]{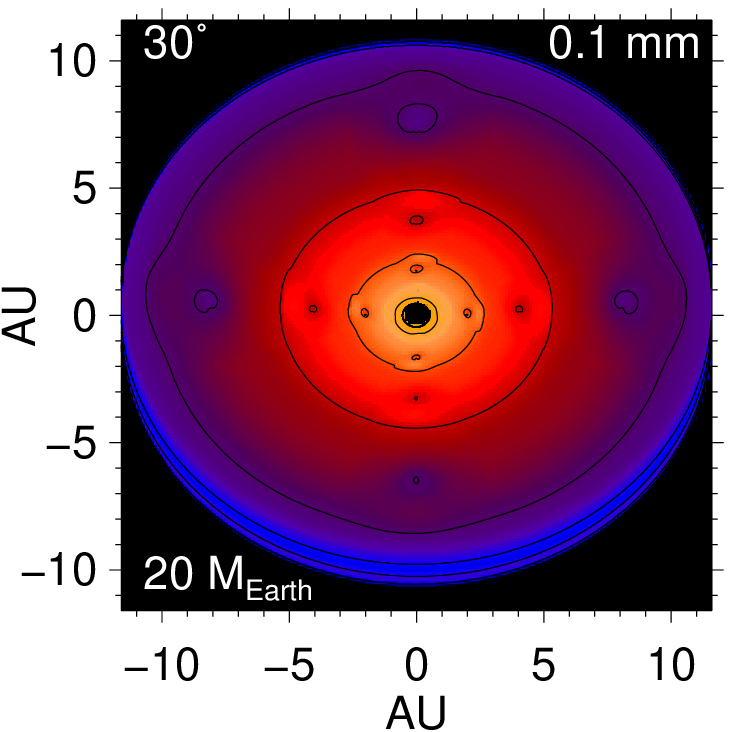}
\includegraphics[width=2.12in]{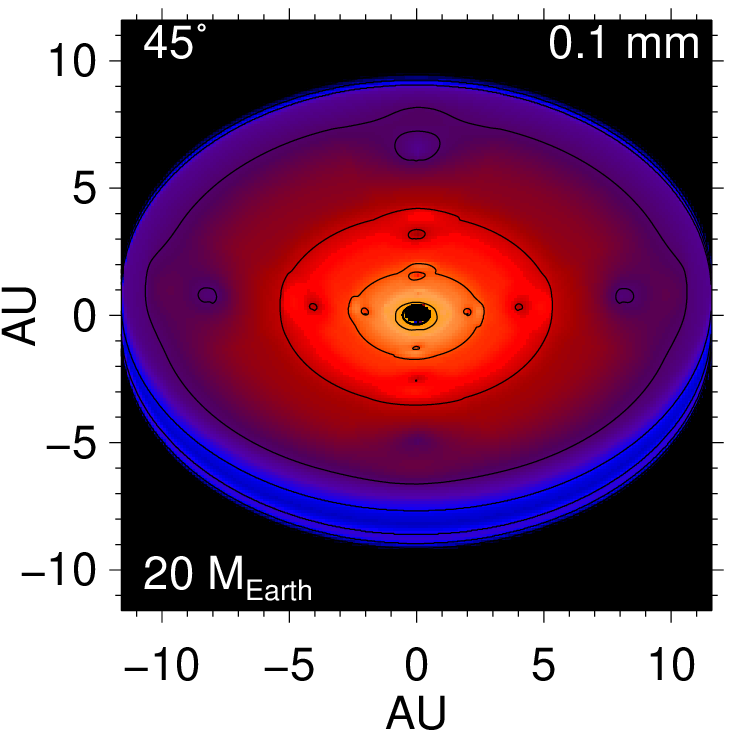}
\includegraphics[width=2.12in]{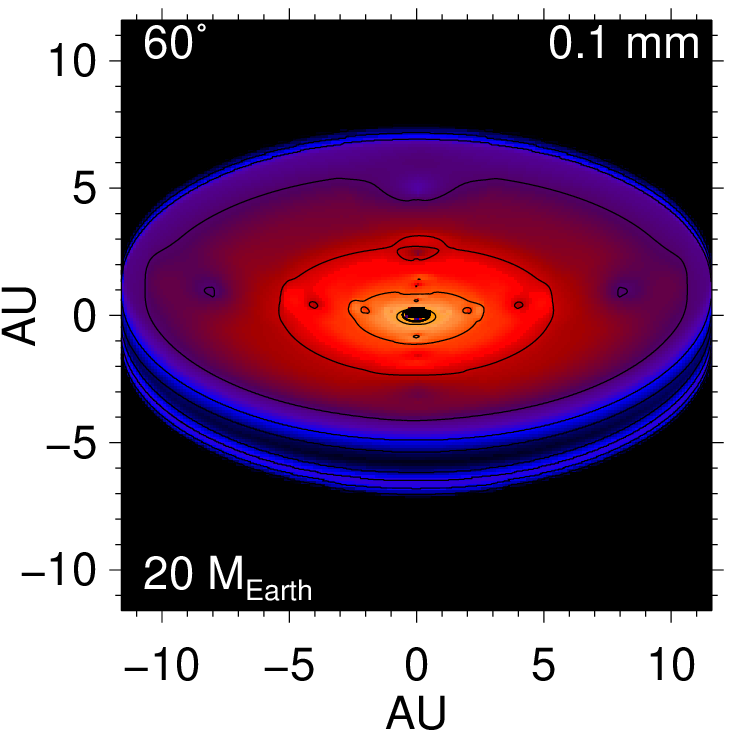}\\
\includegraphics[width=2.12in]{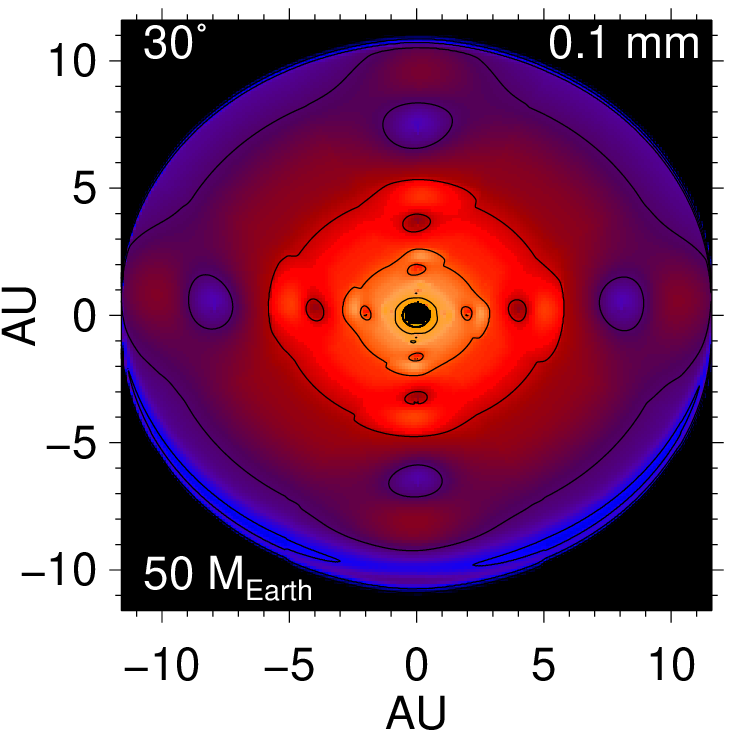}
\includegraphics[width=2.12in]{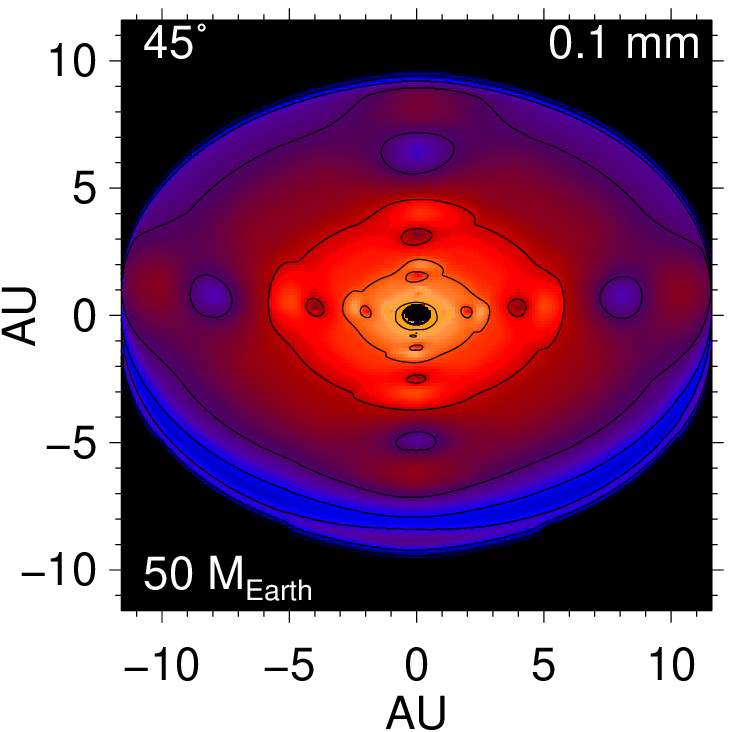}
\includegraphics[width=2.12in]{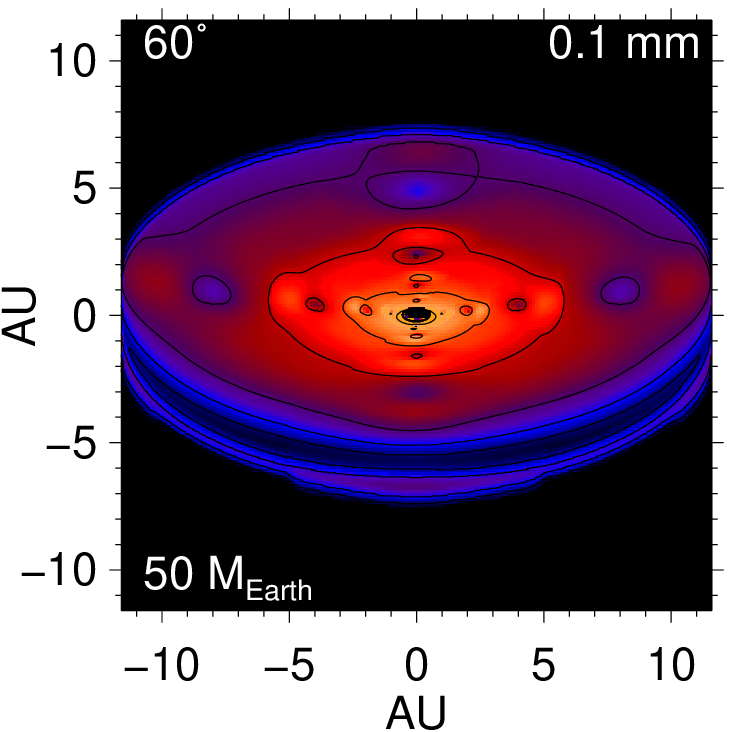}
\caption{\label{hundredmicronincl}
Simulated images of inclined disks perturbed by embedded planets at 
0.1 mm.  See \figref{onemicronincl} for description.  
}
\end{figure}

\begin{figure}
\includegraphics[width=2.12in]{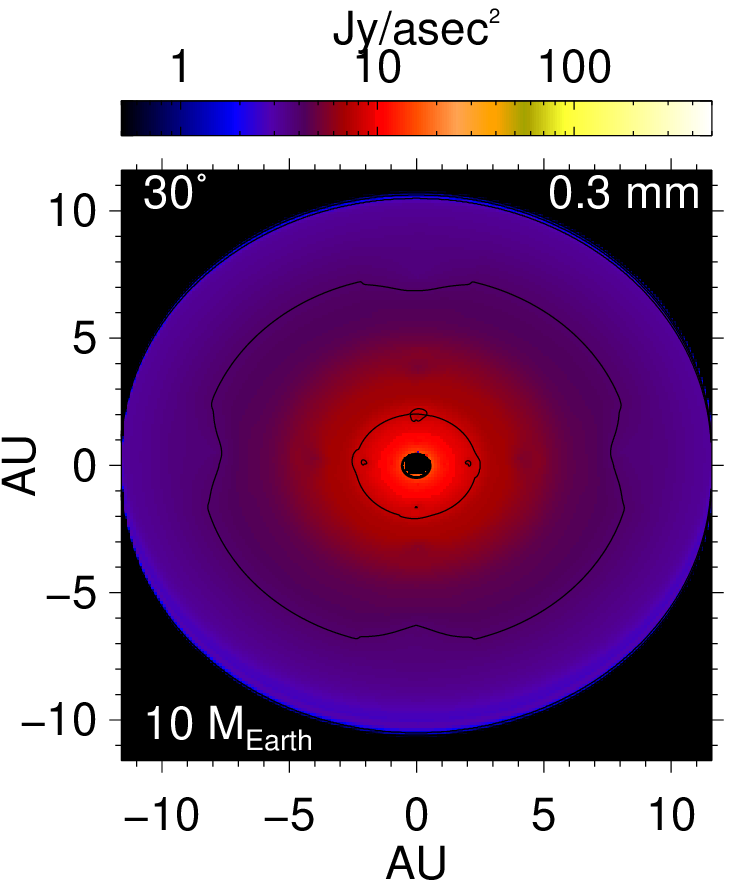}
\includegraphics[width=2.12in]{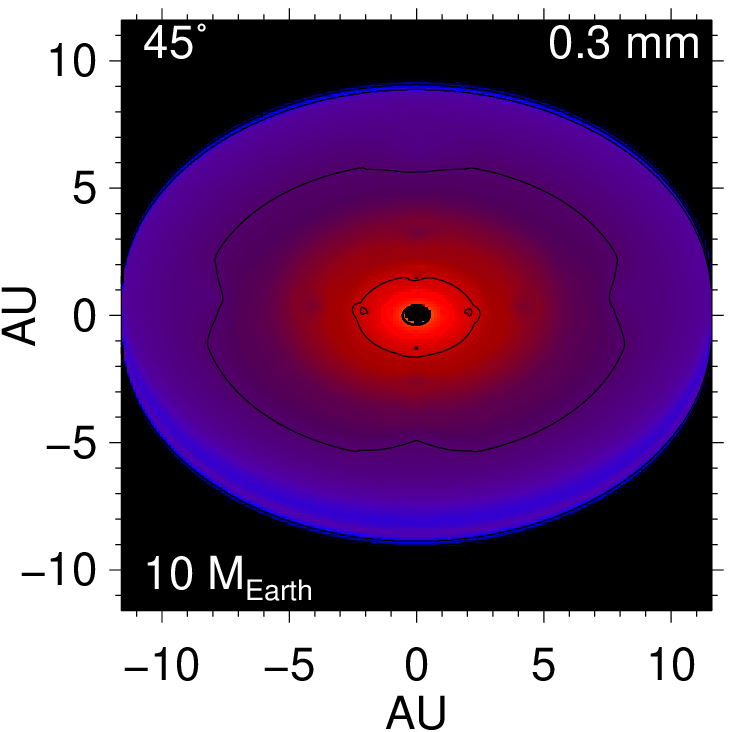}
\includegraphics[width=2.12in]{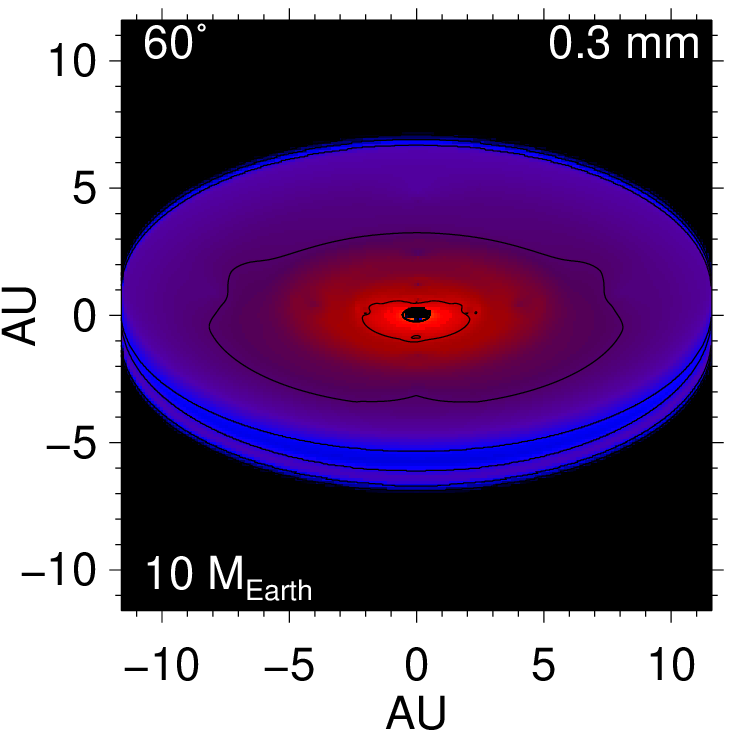}\\
\includegraphics[width=2.12in]{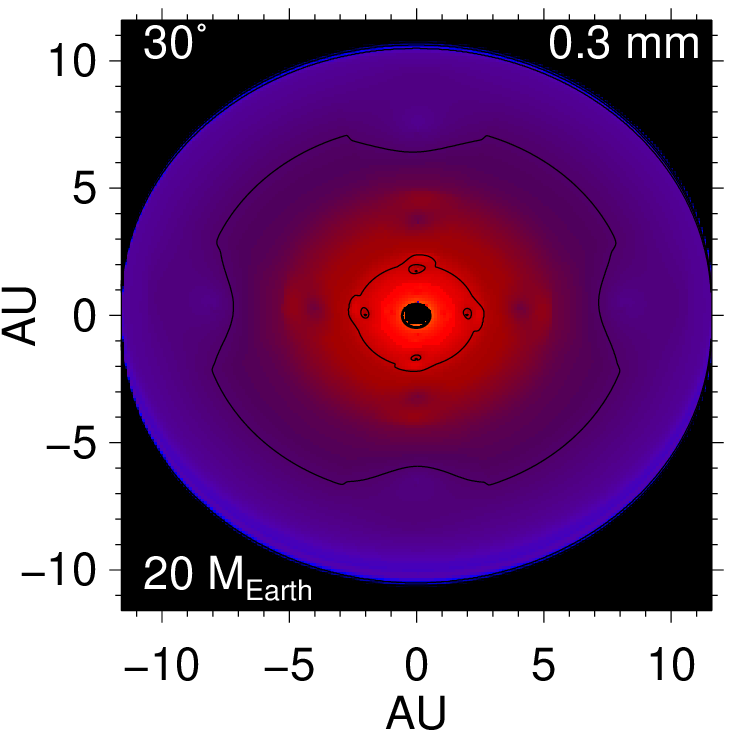}
\includegraphics[width=2.12in]{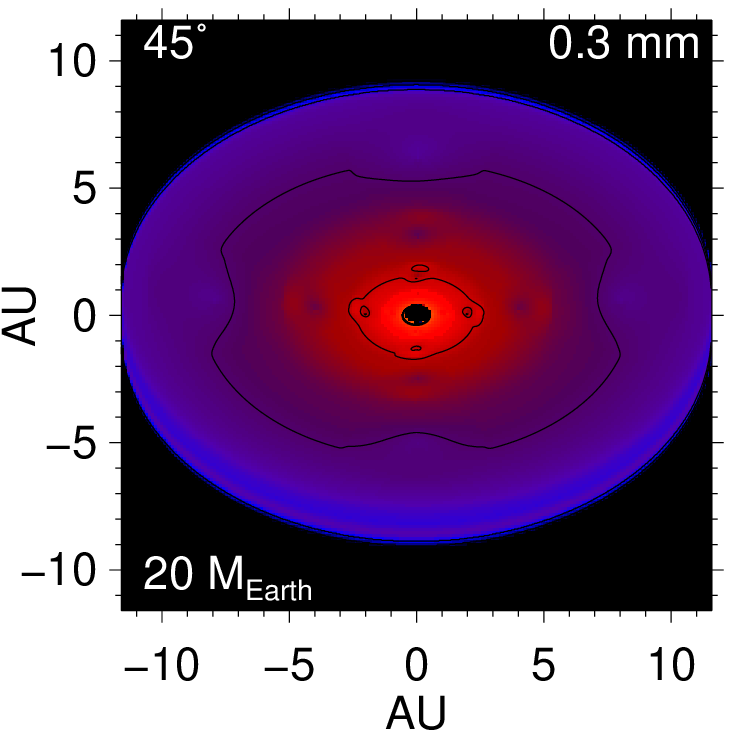}
\includegraphics[width=2.12in]{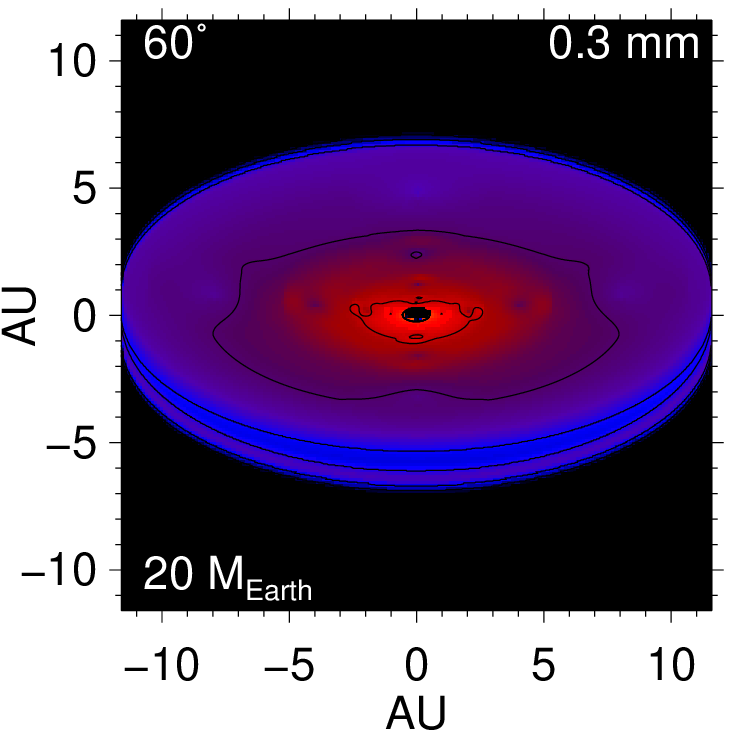}\\
\includegraphics[width=2.12in]{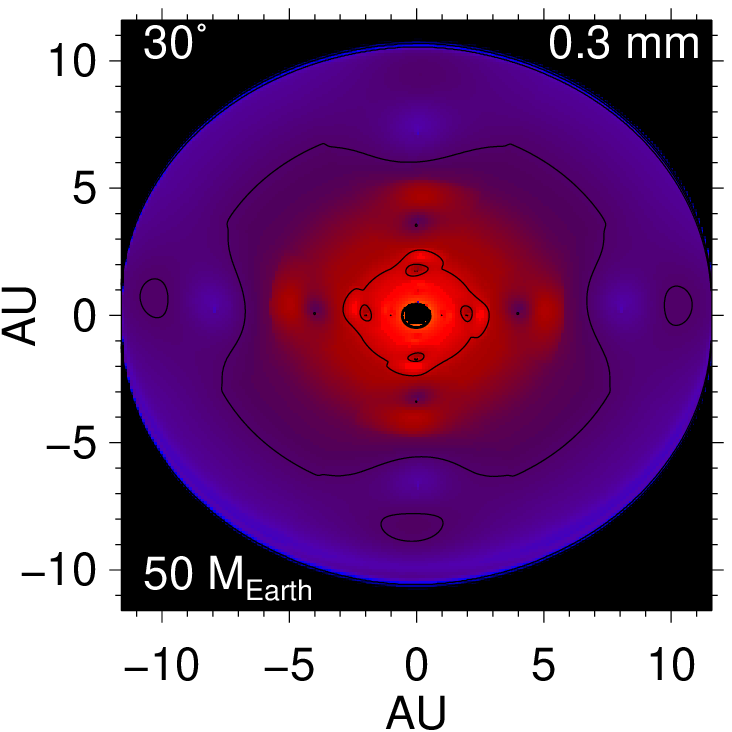}
\includegraphics[width=2.12in]{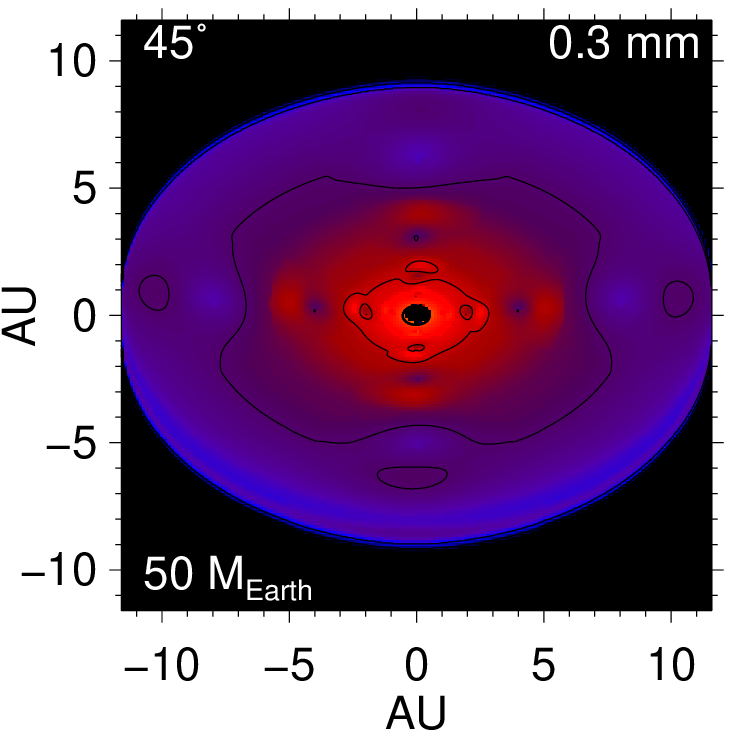}
\includegraphics[width=2.12in]{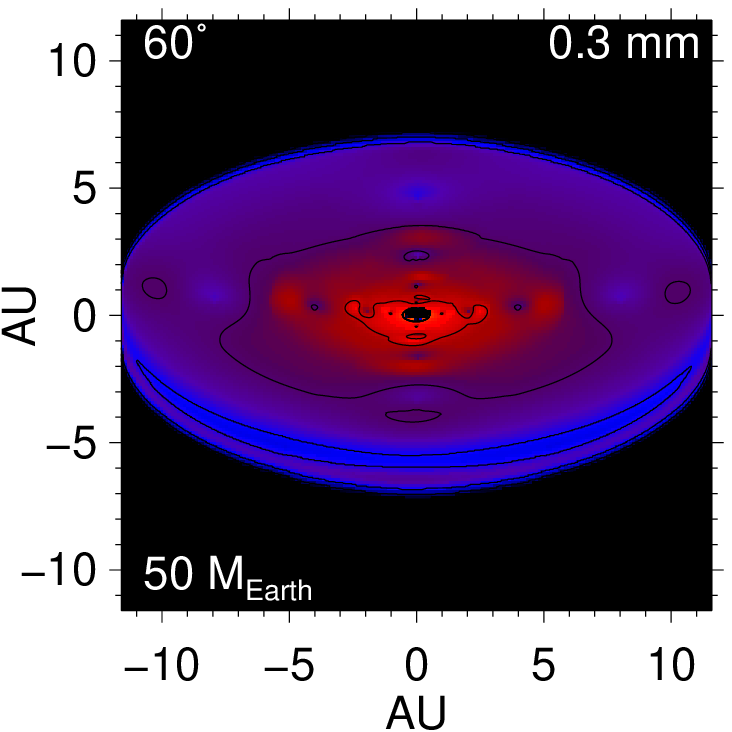}
\caption{\label{threehundredmicronincl}
Simulated images of inclined disks perturbed by embedded planets at 
0.3 mm.  See \figref{onemicronincl} for description.  
}
\end{figure}

In Figures \ref{allfaceon}-\ref{threehundredmicronincl} we show 
the results of simulating images planets in disks from 1 to 300 microns.  
Face-on disks ($0\degr$ inclination) with planets are displayed in 
\figref{allfaceon}.  Each disk image shows a different wavelength, 
with 10, 20 and 50 $M_{\earth}$ planets embedded at 1, 2, 4 and 8 AU.  
The 10 $M_{\earth}$ planets are at the 9 o'clock position 
and the 20 $M_{\earth}$ planets are are the 12 o'clock position.  
The 50 $M_{\earth}$ planets are in the lower right quadrant of the disk.  
They are offset from each other because their simulation boxes 
overlap slightly in the radial direction.  In subsequent figures, 
we allow the boxes to overlap because the perturbation inward 
of the planet is minimal.  

The shadows are roughly centered at the planet's position in radius and 
azimuth, and brightened spots appear on the far side of the 
dimple.  We define the contrast to be $ c= f/f_0-1$ where $f_0$ is 
the radially-dependent 
unperturbed disk brightness and $f$ is the disk brightness in the 
presence of the planet.  
The depths of the shadows/bright spots
regions in the face-on disk 
can then be characterized by the minimum/maximum value of 
the contrast.  
The sizes of the shadows/bright spots are characterized by 
the area enclosed by a contour traced at half the minimum/maximum 
constrast.  
These values are tabulated in columns 5, 6, 8 \& 9 of 
Table \ref{spottable} 
by planet mass (column 1), orbital distance (column 2), and 
wavelength of observation (column 4).  
The ``spot radius'' is the equivalent radius of the half-min/max 
areas $A$,
\begin{equation}
r\sub{spot} = \sqrt{\frac{A}{\pi}}.
\end{equation} 

In \figref{contrast}, we plot the maximum/minimum values of the 
contrast versus wavelength, with line type and color 
indicating orbital distance 
and symbols indicating planet mass.  The contrasts generally increase 
with mass, while the dependence on distance in more complicated.  
The wavelength dependence will be discussed in further detail below.  

\figref{spotradii} shows the dependence of $r\sub{spot}/r\sub{Hill}$ 
with wavelength, with symbol shape and color indicating mass, 
and horizontal offset indicating distance.  The clustering of the points 
indicates that $r\sub{spot}$ scales roughly with $r\sub{Hill}$, 
with brightened spots being generally larger than shadows.  
In other words, 
the size of the shadow scales roughly with the distance from the star
and the cube root of the planet mass.  

In Figures \ref{onemicronincl}, 
\ref{threemicronincl},
\ref{tenmicronincl},
\ref{thirtymicronincl},
\ref{hundredmicronincl} and 
\ref{threehundredmicronincl} we 
show the effect of inclincation on the appearance of the perturbed disk 
at wavelengths of 1, 3, 10, 30, 100, and 300 microns, respectively.  
Each figure shows, from left to right, an angle of inclination 
of $30\degr$, $45\degr$ and $60\degr$.  
From top to bottom, the planet mass is 10, 20, and 50 $M_{\earth}$.  
The inclination axis is always pointing 
upwards, so that the upper part of the disk is further from the 
observer than the bottom part.  In each disk model, planets 
are inserted at 1, 2, 4, and 8 AU, along each minor and major axis 
to demonstrate how the phase of the planet affects the appearance
of the perturbation.

\subsection{Scattered light}

In the optical to near-IR (1 and 3 $\mu$m, 
upper plots in \figref{allfaceon}), 
the image of the disk is predominantly 
from scattered light.  These images trace the contours of the 
surface of the disk.  The shadowing and illumination are clearly visible.  
For reference, the stellar brightness 
is $1.54 (d/100\,\mbox{pc})^{-2}$ Jy at 1 micron, 
and $0.77 (d/100\,\mbox{pc})^{-2}$ Jy at 3 microns, 
where $d$ is the distance to the star.  
At these wavelengths, the brightness of the star overwhelms 
any emission from the disk, so very good starlight suppression, 
i.e.~coronography, is necessary to resolve any features in the disk.  
Coronography is also a good way to suppress 
emission from hot inner walls of the disk, since this 
may be a significant source of near-IR flux in the inner 0.5 AU of the disk.

The shadow contrast is generally deeper than the bright contrast 
because it is easier to create a depression 
than it is to puff up disk material due to heating, 
especially in the surface layers of the disk where the 
density is very low.  
As seen in \citet{HJC_model}, the heating from 
illumination is greater interior to 8 AU, because the 
densities are higher in the inner regions.  
The exception to this is for the 
largest planet at 2-4 AU, where the bright contrast 
rises above 1, whereas the shadow contrast has a floor 
at $-1$.  On the other hand, the area of the brightened regions 
are generally larger than the shadowed region, 
except for the smallest planet at 8 AU, where 
the amount of brightening is very small.  

The effect of inclination on planet dimples can be seen in 
Figures \ref{onemicronincl} and \ref{threemicronincl}.  
One general effect is that as the angle of inclination increases, 
the brightness of the disk also increases.  
Examining equation (\ref{iscatt}), we see that as $i$, the angle of 
observation with respect to the surface normal, increases, $\cos i$ 
decreases, increasing the overall value of $I^{s}$. 
The interpretation for this is that the observer is viewing 
more scatterers along the line of sight through the disk, 
thus the disk appears brighter.  
Since the disk images are tipped up so that the bottom half 
is closer to the observer, the value of $i$ is greater, so the 
bottom halves appear brighter.  The planet perturbation on top 
of this general effect causes the rebrightened region of the 
disk to appear brighter still when it is on the near side.

\subsection{Mid-IR: thermal emission from the surface}

The mid-IR (10 and 30 $\mu$m, center plots in \figref{allfaceon})
probes thermal emission from the surface layers of the disk.
At these wavelengths, the observed features are caused by 
temperature perturbations resulting from cooling in the shadowed 
regions and heating in the illuminated regions.  
The size and contrast of the perturbations show the same variation 
with planet size and distance as seen in the scattered light 
images.  Again, the shadows appear deeper than the brightened spots,  
except when the bright contrast $\gtrsim 1.8$ because of the natural 
floor in the shadow contrast at $-1$.  
The stellar brightness is 
$0.107 (d/100\,\mbox{pc})^{-2}$ Jy at 10 microns, 
and $1.34\times10^{-2} (d/100\,\mbox{pc})^{-2}$ Jy at 30 microns. 

The greatest amount of contrast occurs at 10 $\mu$m.  
The 10 $\mu$m image also shows the most radial contrast 
in the disk overall because the outer regions of the disk are 
too cold to emit effectively at this wavelength: 10 $\mu$m is 
shortward of the exponential cutoff in thermal emission in the 
outer disk.  
The steep background radial gradient of the unperturbed disk 
itself will make it difficult to isolate perturbations from embedded planets.  

In the inclined disks (Figures \ref{tenmicronincl} and \ref{thirtymicronincl})
there appears to be a brightening associated with higher inclinations, 
similar to the scattered light images, but of lesser magnitude.  
This brightening is caused by an optical depth effect, 
because as more heated or cooled material aligns along the line of 
sight, the stronger or weaker the thermal emission becomes.

\subsection{Far-IR to radio: thermal emission from the midplane}

At sub-millimeter wavelengths, the stellar brightness is negligible 
compared to the thermal emission from the disk.  
In this regime the perturbations induced 
by the embedded planets are hardly visible 
(0.1 mm and 0.3 mm, lower plots in \figref{allfaceon}).  
The contrast is modest compared to shorter wavelengths.  
The scatter in $r\sub{spot}$ at 300 $\mu$m in 
\figref{spotradii} is largely because the contrast is so modest 
that the half-min/max contours are hard to constrain.  

The low contrast arises because the thermal structure of the disk 
is affected primarily in the upper layers of the disk.  As the disk 
becomes optically thick to its own thermal radiation, the effects 
of small perturbations at the surface become washed out.  
The contrast between the inner and outer disk also becomes 
less pronounced because these wavelengths are in the 
Rayleigh-Jeans regime of the blackbody spectra.  

However, other effects not modeled in this study may very 
well create observable signatures.  In particular, dynamics within 
the Hill sphere is not properly included.  Circumplanetary disk 
formation and accretion onto the forming planet will create 
a density enhancement and heating at the midplane, which will 
increase the thermal emission in this region, particularly 
at wavelengths where the disk becomes optically thin.  Better 
models need to be carried out to properly model planet signatures 
in the radio, particularly to determine whether small planets 
create shadows versus brightenings.

\section{Discussion: Observability}\label{discussion}

In order to observe the phenomena modeled in this paper, we need 
to look to nearby star-forming regions to find sufficiently young 
protoplanetary disks.  The Taurus-Auriga star-forming is one 
of the nearest, at about 140 pc \citep{1994KenyonDobrzyckaHartmann}.  
At this distance, a 1 AU perturbation subtends 0.007\arcsec.  

% lambda/D = 1/140 arcsec = 3.463e-8 radians
% D = lambda/3.5e-8 = (lambda/microns)/3.5e-2 meters
%   = (lambda/microns)* 28.877 m

The best wavelengths for observing planet shadows as modeled in this 
paper are in the visible to mid-infrared.  Scattered light 
images are sensitive to small perturbations in density 
at the surface of the disk.  However, this is both an advantage and 
disadvantage.  Small variations such as those caused by small planets 
can cause large shadowing effects.  On the other hand, density 
variations might be caused by other processes, such as 
outflows, disk turbulence, and dust clumpiness, just to name a 
few.  Differentiating planet-induced shadows from those caused 
by other processes may prove to be difficult.  

At 1 $\mu$m, a diffraction limit of $\lambda/D=7$ mas 
requires a telescope diameter of 29 meters.  
This is larger than any current ground-based telescope or 
planned space telescope.  However, 
with the development of large segmented-mirror telescopes 
such as the Giant Magellan Telescope (GMT)\footnote{http://www.gmto.org/}
and the Thirty Meter Telescope (TMT)\footnote{http://www.tmt.org/}, 
these surface shadows may be resolvable in the near future.  
However, scattered light observations of the planet perturbations 
requires very good suppression of the starlight at small 
inner working angles.  

For mas resolution at longer wavelengths, 
single-dish telescopes are increasingly infeasible, but 
interferometry is a promising alternative.  
The Center of High Angular Resolution Astronomy (CHARA) 
Array\footnote{http://www.chara.gsu.edu/CHARA/index.php} 
at Mt.~Wilson Observatory has baselines of up to 330 meters 
and operates in the optical and $2.0-2.5\,\mu$m range, 
allowing for sub-mas resolution.  
The Very Large Telescope Interferometer 
(VLTI)\footnote{http://www.eso.org/sci/facilities/paranal/telescopes/vlti/index.html} can achieve up to 2 mas resolution in J, H and K bands ($1-2.4\,\mu$m)
with the AMBER instrument.  VLTI-MIDI is a two-way beam 
combiner at N-band ($8-13\,\mu$m), for an angular resolution of about 10 mas. 
A next-generation mid-IR instrument for VLTI, 
MATISSE\footnote{http://www.oca.eu/matisse/}, is also under development 
for L, N, M, and Q bands ($3-25\,\mu$m).  
The challenge for these instruments is covering sufficient $u$-$v$ 
space to build up an image, 
achieving sufficient dynamic resolution so that the star does
not overwhelm the disk emission, and having enough sensitivity 
to detect the planet shadows.  

Mid-infrared wavelengths, which probe temperature 
perturbations caused by shadowing and illumination at the 
surface, are better for identifying planet perturbations.  
This is because shadows and brightenings caused by transitory 
phenomena will not persist as long those caused by an 
embedded planet, and will not result in strong cooling and 
heating in the surface layers.  Although the shadow contrast 
is good at 10 $\mu$m, the dynamic range between the inner and 
outer disk is quite high, over 4 orders of magnitude from 0.5 to 10 AU.  
Dynamic range is less of an issue at 30 $\mu$m.  
The 30 $\mu$m images 
shown in Figures \ref{allfaceon} and \ref{thirtymicronincl} 
are very similar to the 1 $\mu$m images in Figures \ref{allfaceon}
and \ref{onemicronincl}: these two different wavelengths 
trace the planet perturbations equally well.  

A resolving power of 7 mas at 30 $\mu$m 
requires firstly getting above the Earth's atmosphere, 
then having a baseline of 870 m: this would almost 
certainly require a formation-flying space mission. 
However, there are no near-term prospects for such a mission.  

In the radio, the Atacama Large Millimeter/submillimeter Array 
(ALMA)\footnote{http://almaobservatory.org} will have 
mas resolution when it reaches its full operating 
capacity in 2012.  While the resolution would be 
good enough to resolve a planet shadow, the perturbations 
have very low contrast at 0.3 mm, the shortest wavelength
that ALMA will operate at.  The problem here is one of 
sensitivity rather than angular resolution.  

This study was restricted to planets forming interior to 10 AU, 
based on scenarios of giant planet formation within our 
own Solar System.  
However, discoveries of planets around HR 8799 and Fomalhaut 
indicate that exoplanets can form tens to hundreds of AU 
from their stars \citep{2008HR8799,2008Fomalhaut}. 
The size of planet shadows in protoplanetary disks scale 
roughly with the planet-star distance. 
So there is hope that planets forming at these far distances 
from their stars might be observable in the near future, 
perhaps even with current instrumentation.  

\section{Conclusions}\label{conclusion}

Observable perturbations from planets roughly scale with 
planet mass and orbital distance.  The features are most 
easily observed in visible to mid-infrared wavelengths.  
The visible regime probes scattered light off the surface 
of the disk, revealing shadows caused by the gravitational 
potential of the planet.  Mid-IR wavelengths reveal cooling/heating 
from shadowed/brightened regions at the surface.  
Confirmation of shadows and brightenings seen in scattered light 
images of disks should therefore be carried out at near- to mid-IR 
wavelengths.  In longer wavelengths toward the sub-millimeter, 
the contrast of the perturbations becomes smaller as the disk 
becomes optically thin.  

The validity of the simulations presented in this study is limited 
by the fact that hydrodynamics are not included.  
This means that large-scale effects such as gap-clearing 
are not well-modeled.  If an annular gap does form, this density 
perturbation will be subjected to shadowing and illumination 
in the same way local dimples are, so radiative feedback on 
gaps should enhance these structures as well.  Thus, the 
local planet shadows modeled in this paper represent lower limits 
on perturbations to disks that might be caused by embedded 
plaent cores.  

Another missing physical process is accretion onto the 
forming planet core.  This process is difficult to capture even 
in high resolution three-dimensional hydrodynamic simulations because 
of the extremely high dynamic range necessary to resolve the final 
capture of gas into the atmosphere of the growing planet.  
Heating from accretion onto the planet could alter the results 
of this study.  Simulations by \citet{2005Hubickyj_etal} 
indicate that growing planets accrete slowly over a long period 
of time, but once it crosses a threshold mass of roughly 
$10-20\,M_{\earth}$, 
it accretes gas rapidly over a short time period.  
Thus, accretional heating is likely to be only 
transitionally important.  On the other hand, planets might grow so 
rapidly that 50 Earth mass planets might never be seen, 
being only a transitional stage on the way to becoming a 
Jovian-type planet.  

\acknowledgements
The author thanks 
John Debes for helpful discussions and advice in 
developing the dust opacity model used in this paper; 
Lee Mundy and an anonymous referee for helpful comments; 
and Matthew Condell for assistance with software development. 
This work was performed under contract with the Jet
Propulsion Laboratory (JPL) funded by NASA through the Michelson
Fellowship Program. JPL is managed for NASA by the California
Institute of Technology.

\bibliographystyle{apj}
\bibliography{}

\input{table1.tex}

\end{document}

%% file: table1.tex
\begin{deluxetable}{cclrllrllr}
\tablecaption{\label{spottable}Contrast and size of shadowed and brightened 
regions.  }
\tablehead{
\colhead{$m_p$} &
\colhead{$a$} &
\colhead{$r\sub{Hill}$} &
\colhead{$\lambda$} &
\colhead{shadow } &
\colhead{area} &
\colhead{} &
\colhead{bright } &
\colhead{area} &
\colhead{}\\
\colhead{($M_{\earth}$)} &
\colhead{(AU)} &
\colhead{(AU)} &
\colhead{($\mu$m)} &
\colhead{contrast} &
\colhead{(AU$^2$)} &
\colhead{\raisebox{1.5ex}[0pt]{$\displaystyle{\frac{r\sub{spot}}{r\sub{Hill}}}$}} &
\colhead{contrast} &
\colhead{(AU$^2$)} &
\colhead{\raisebox{1.5ex}[0pt]{$\displaystyle{\frac{r\sub{spot}}{r\sub{Hill}}}$}}
}
\startdata
  10 &  1 & 0.0216 &   1 & -0.564 & 0.0118 &   5.70 &  0.143 & 0.0239 &   8.09 \\
  10 &  1 & 0.0216 &   3 & -0.578 & 0.0111 &   5.53 &  0.138 & 0.0252 &   8.31 \\
  10 &  1 & 0.0216 &  10 & -0.717 &0.00950 &   5.10 &  0.306 & 0.0294 &   8.98 \\
  10 &  1 & 0.0216 &  30 & -0.270 & 0.0209 &   7.56 &  0.129 & 0.0323 &   9.41 \\
  10 &  1 & 0.0216 & 100 & -0.116 & 0.0258 &   8.41 & 0.0539 & 0.0329 &   9.49 \\
  10 &  1 & 0.0216 & 300 &-0.0687 & 0.0301 &   9.08 & 0.0333 & 0.0327 &   9.47 \\
  10 &  2 & 0.0431 &   1 & -0.643 & 0.0519 &   5.97 &  0.295 &  0.106 &   8.54 \\
  10 &  2 & 0.0431 &   3 & -0.634 & 0.0527 &   6.01 &  0.280 &  0.105 &   8.50 \\
  10 &  2 & 0.0431 &  10 & -0.864 & 0.0267 &   4.28 &  0.548 &  0.115 &   8.88 \\
  10 &  2 & 0.0431 &  30 & -0.395 & 0.0858 &   7.67 &  0.303 &  0.124 &   9.21 \\
  10 &  2 & 0.0431 & 100 & -0.154 &  0.111 &   8.71 &  0.107 &  0.128 &   9.38 \\
  10 &  2 & 0.0431 & 300 &-0.0886 &  0.130 &   9.45 & 0.0601 &  0.130 &   9.43 \\
  10 &  4 & 0.0862 &   1 & -0.610 &  0.247 &   6.50 &  0.182 &  0.287 &   7.02 \\
  10 &  4 & 0.0862 &   3 & -0.604 &  0.251 &   6.56 &  0.171 &  0.283 &   6.97 \\
  10 &  4 & 0.0862 &  10 & -0.827 &  0.118 &   4.49 &  0.413 &  0.366 &   7.92 \\
  10 &  4 & 0.0862 &  30 & -0.553 &  0.292 &   7.07 &  0.250 &  0.368 &   7.94 \\
  10 &  4 & 0.0862 & 100 & -0.178 &  0.513 &   9.38 & 0.0709 &  0.356 &   7.81 \\
  10 &  4 & 0.0862 & 300 &-0.0953 &  0.589 &   10.0 & 0.0320 &  0.338 &   7.61 \\
  10 &  8 &  0.172 &   1 & -0.466 &   1.01 &   6.59 & 0.0346 &  0.420 &   4.24 \\
  10 &  8 &  0.172 &   3 & -0.464 &   1.02 &   6.62 & 0.0352 &  0.292 &   3.54 \\
  10 &  8 &  0.172 &  10 & -0.651 &  0.394 &   4.11 &  0.162 &  0.635 &   5.21 \\
  10 &  8 &  0.172 &  30 & -0.571 &  0.671 &   5.36 &  0.118 &  0.686 &   5.42 \\
  10 &  8 &  0.172 & 100 & -0.165 &   2.28 &   9.87 & 0.0121 &  0.281 &   3.47 \\
  10 &  8 &  0.172 & 300 &-0.0725 &   2.97 &   11.3&0.000450 &   3.92 &   13.0 \\
  20 &  1 & 0.0272 &   1 & -0.696 & 0.0194 &   5.79 &  0.302 & 0.0463 &   8.94 \\
  20 &  1 & 0.0272 &   3 & -0.704 & 0.0189 &   5.71 &  0.301 & 0.0481 &   9.12 \\
  20 &  1 & 0.0272 &  10 & -0.739 & 0.0216 &   6.10 &  0.637 & 0.0525 &   9.52 \\
  20 &  1 & 0.0272 &  30 & -0.336 & 0.0579 &   10.0 &  0.271 &  0.106 &   13.5 \\
  20 &  1 & 0.0272 & 100 & -0.153 & 0.0358 &   7.86 &  0.113 & 0.0589 &   10.1 \\
  20 &  1 & 0.0272 & 300 &-0.0924 & 0.0404 &   8.35 & 0.0716 & 0.0590 &   10.1 \\
  20 &  2 & 0.0543 &   1 & -0.826 & 0.0889 &   6.20 &  0.790 &  0.201 &   9.33 \\
  20 &  2 & 0.0543 &   3 & -0.814 & 0.0901 &   6.24 &  0.765 &  0.203 &   9.35 \\
  20 &  2 & 0.0543 &  10 & -0.932 & 0.0758 &   5.72 &   1.37 &  0.213 &   9.58 \\
  20 &  2 & 0.0543 &  30 & -0.535 &  0.129 &   7.47 &  0.824 &  0.233 &   10.0 \\
  20 &  2 & 0.0543 & 100 & -0.234 &  0.141 &   7.80 &  0.285 &  0.256 &   10.5 \\
  20 &  2 & 0.0543 & 300 & -0.227 & 0.0632 &   5.22 &  0.167 &  0.263 &   10.7 \\
  20 &  4 &  0.109 &   1 & -0.835 &  0.436 &   6.86 &  0.601 &  0.778 &   9.16 \\
  20 &  4 &  0.109 &   3 & -0.827 &  0.442 &   6.91 &  0.581 &  0.775 &   9.15 \\
  20 &  4 &  0.109 &  10 & -0.966 &  0.327 &   5.94 &   1.13 &  0.835 &   9.49 \\
  20 &  4 &  0.109 &  30 & -0.673 &  0.650 &   8.37 &  0.846 &  0.815 &   9.38 \\
  20 &  4 &  0.109 & 100 & -0.284 &  0.734 &   8.90 &  0.256 &  0.894 &   9.82 \\
  20 &  4 &  0.109 & 300 & -0.275 &  0.228 &   4.97 &  0.128 &  0.915 &   9.94 \\
  20 &  8 &  0.217 &   1 & -0.727 &   1.74 &   6.86 &  0.208 &   1.88 &   7.13 \\
  20 &  8 &  0.217 &   3 & -0.720 &   1.78 &   6.92 &  0.192 &   1.83 &   7.02 \\
  20 &  8 &  0.217 &  10 & -0.882 &   1.09 &   5.41 &  0.433 &   2.41 &   8.06 \\
  20 &  8 &  0.217 &  30 & -0.803 &   1.66 &   6.70 &  0.404 &   2.32 &   7.91 \\
  20 &  8 &  0.217 & 100 & -0.290 &   3.36 &   9.52 &  0.105 &   2.11 &   7.55 \\
  20 &  8 &  0.217 & 300 & -0.224 &   1.24 &   5.78 & 0.0384 &   1.92 &   7.20 \\
  50 &  1 & 0.0369 &   1 & -0.790 & 0.0410 &   6.20 &  0.657 & 0.0743 &   8.35 \\
  50 &  1 & 0.0369 &   3 & -0.797 & 0.0404 &   6.15 &  0.666 & 0.0775 &   8.52 \\
  50 &  1 & 0.0369 &  10 & -0.819 & 0.0441 &   6.43 &   1.36 &  0.120 &   10.6 \\
  50 &  1 & 0.0369 &  30 & -0.537 & 0.0394 &   6.07 &  0.571 &  0.102 &   9.80 \\
  50 &  1 & 0.0369 & 100 & -0.209 & 0.0545 &   7.14 &  0.232 &  0.104 &   9.87 \\
  50 &  1 & 0.0369 & 300 & -0.131 & 0.0582 &   7.39 &  0.149 &  0.105 &   9.94 \\
  50 &  2 & 0.0737 &   1 & -0.914 &  0.196 &   6.78 &   1.64 &  0.276 &   8.04 \\
  50 &  2 & 0.0737 &   3 & -0.910 &  0.196 &   6.77 &   1.60 &  0.281 &   8.12 \\
  50 &  2 & 0.0737 &  10 & -0.957 &  0.189 &   6.66 &   3.27 &  0.267 &   7.92 \\
  50 &  2 & 0.0737 &  30 & -0.731 &  0.204 &   6.91 &   1.77 &  0.332 &   8.83 \\
  50 &  2 & 0.0737 & 100 & -0.333 &  0.220 &   7.18 &  0.557 &  0.403 &   9.72 \\
  50 &  2 & 0.0737 & 300 & -0.211 &  0.229 &   7.32 &  0.328 &  0.428 &   10.0 \\
  50 &  4 &  0.147 &   1 & -0.964 &  0.885 &   7.20 &   1.65 &   1.40 &   9.05 \\
  50 &  4 &  0.147 &   3 & -0.963 &  0.883 &   7.19 &   1.62 &   1.42 &   9.12 \\
  50 &  4 &  0.147 &  10 & -0.998 &  0.843 &   7.03 &   3.46 &   1.61 &   9.71 \\
  50 &  4 &  0.147 &  30 & -0.864 &   1.02 &   7.71 &   2.57 &   1.57 &   9.58 \\
  50 &  4 &  0.147 & 100 & -0.422 &  0.986 &   7.60 &  0.683 &   1.94 &   10.7 \\
  50 &  4 &  0.147 & 300 & -0.253 &  0.997 &   7.64 &  0.345 &   2.03 &   10.9 \\
  50 &  8 &  0.295 &   1 & -0.979 &   3.00 &   6.63 &  0.837 &   5.53 &   9.00 \\
  50 &  8 &  0.295 &   3 & -0.971 &   3.04 &   6.68 &  0.810 &   5.53 &   9.00 \\
  50 &  8 &  0.295 &  10 & -0.985 &   2.92 &   6.54 &   1.53 &   6.13 &   9.48 \\
  50 &  8 &  0.295 &  30 & -0.942 &   3.51 &   7.17 &   1.60 &   5.98 &   9.36 \\
  50 &  8 &  0.295 & 100 & -0.436 &   4.35 &   7.98 &  0.486 &   6.51 &   9.77 \\
  50 &  8 &  0.295 & 300 & -0.274 &   3.42 &   7.08 &  0.202 &   6.77 &   9.96 
\enddata
\end{deluxetable}